\documentclass[11pt]{article}

\usepackage{verbatim}
\usepackage{amsmath}
\usepackage{amsfonts}
\usepackage{amssymb}
\usepackage{bbm}
\usepackage{latexsym}
\usepackage{lscape} 
\usepackage{epsfig}
\usepackage{color}
\usepackage{graphicx}
\usepackage{subfigure}
\usepackage[all]{xy}
\usepackage{pbox}
\usepackage{multirow}

\usepackage{hyperref}

\usepackage{multirow}
\usepackage{mathrsfs}

\usepackage{amscd}
\usepackage{cite}

\usepackage{tikz}

\renewcommand{\d}{\mathrm{d}}

\definecolor{gray}{gray}{.5}

\DeclareMathSymbol{\mg}{\mathrel}{symbols}{"1D}

\newcommand{\ga}{\alpha}
\newcommand{\gb}{\beta}
\renewcommand{\gg}{\gamma}
\newcommand{\gd}{\delta}
\renewcommand{\ge}{\epsilon}
\newcommand{\gve}{\varepsilon}
\newcommand{\gf}{\phi}
\newcommand{\gvf}{\varphi}

\newcommand{\gx}{\xi}
\newcommand{\gm}{\mu}
\newcommand{\gn}{\nu}

\newcommand{\gk}{\kappa}
\newcommand{\gl}{\lambda}
\newcommand{\gr}{\rho}

\newcommand{\gs}{\sigma}

\newcommand{\go}{\omega}

\newcommand{\gp}{\pi}
\newcommand{\gps}{\psi}
\newcommand{\get}{\eta}

\newcommand{\gG}{\Gamma}
\newcommand{\gD}{\Delta}
\newcommand{\gF}{\Phi}

\newcommand{\cA}{{\cal A}}

\newcommand{\cC}{{\cal C}}
\newcommand{\cD}{{\cal D}}
\newcommand{\cE}{{\cal E}}
\newcommand{\cF}{{\cal F}}
\newcommand{\cG}{{\cal G}}
\newcommand{\cH}{{\cal H}}

\newcommand{\cK}{{\cal K}}
\newcommand{\cL}{{\cal L}}
\newcommand{\cM}{{\cal M}}

\newcommand{\cP}{{\cal P}}
\newcommand{\cQ}{{\cal Q}}
\newcommand{\cR}{{\cal R}}

\newcommand{\cU}{{\cal U}}

\newcommand{\cW}{{\cal W}}

\newcommand{\cZ}{{\cal Z}}
\newcommand{\tV}{{\widetilde V}}

\newcommand{\tX}{{\widetilde X}}

\newcommand{\Tr}{\mbox{Tr}}

\newcommand{\Id}{\text{\small 1}\hspace{-3.5pt}\text{1}}

\newcommand{\ra}{\rightarrow}

\newcommand{\der}{\partial}

\newcommand{\dsp}{\displaystyle}

\newcommand{\beq}{\begin{equation}}
\newcommand{\eeq}{\end{equation}}
\newcommand{\barr}{\begin{array}}
\newcommand{\earr}{\end{array}}
\newcommand{\equ}[1]{\begin{gather} #1 \end{gather}}
\newcommand{\equa}[1]{\begin{align} #1 \end{align}}

\newcommand{\enums}[1]{\begin{enumerate} #1 \end{enumerate}}
\newcommand{\tabu}[2]{\begin{tabular}{#1} #2 \end{tabular}}
\newcommand{\arry}[2]{\begin{array}{#1} #2 \end{array}}

\newcommand{\pmtrx}[1]{\begin{pmatrix} #1 \end{pmatrix}}

\newcommand{\non}{\nonumber}

\newcounter{oldcounter}

\newcommand{\bgm}{{\bar\mu}}

\newcommand{\bgps}{{\bar\psi}}

\newcommand{\tgg}{{\tilde \gamma}}

\newcommand{\ba}[2]{\[\begin{array}{#2}\label{#1}}
\newcommand{\ea}{\end{array}\]}
\newcommand{\be}{\begin{equation}}
\newcommand{\ee}{\end{equation}}
\newcommand{\bea}{\begin{eqnarray}}
\newcommand{\eea}{\end{eqnarray}}

\newcommand{\dl}{\partial_L}
\newcommand{\dr}{\partial_R}
\newcommand{\wt}[1]{\widetilde{#1}}

\newcommand{\vr}{V_R}

\renewcommand{\d}{\textrm{d}}
\newcommand{\dsigma}{\d^2 \sigma}
\newcommand{\dpp}{\frac{\d^d p}{(2 \pi)^d \gm^{d-2}}}

\newcommand{\G}{\mathcal{G}}

\begin{document}

\thispagestyle{empty}

\begin{flushright}
LMU-ASC 58/13 \\
MPP-2013-239  \\
TUM-HEP 904/13
\end{flushright}
\vskip 2 cm
\begin{center}
{\Large {\bf 
Renormalization of a Lorentz invariant doubled worldsheet theory 
} 
}
\\[0pt]

\bigskip
\bigskip
\bigskip {\large
{\bf Stefan Groot Nibbelink$^{a,}$}\footnote{
E-mail: Groot.Nibbelink@physik.uni-muenchen.de},
{\bf Florian Kurz$^{a,b,}$}\footnote{
E-mail: Florian.Kurz@physik.uni-muenchen.de},
{\bf Peter Patalong$^{a,c,}$}\footnote{
E-mail: Peter.Patalong@physik.uni-muenchen.de}
\bigskip }\\[0pt]
\vspace{0.23cm}
${}^a$ {\it 
Arnold Sommerfeld Center for Theoretical Physics,\\
~~Ludwig-Maximilians-Universit\"at M\"unchen, 80333 M\"unchen, Germany
 \\[1ex]} 
${}^b$ {\it 
Physik Department, Technische Universit\"at M\"unchen, \\ 
James--Franck--Stra\ss e 1, 85748 Garching, Germany
\\[1ex]} 
${}^c$ {\it  
Max--Planck--Institut f\"ur Physik, F\"ohringer Ring 6, 80805 M\"unchen, Germany
}
\\[1ex] 
\bigskip
\end{center}

\subsection*{\centering Abstract}

Manifestly T--duality covariant worldsheet string models can be constructed by doubling the coordinate fields.
We describe the underlying gauge symmetry of a recently proposed Lorentz invariant doubled worldsheet theory that makes half of the worldsheet degrees of freedom redundant. 
By shifting the Lagrange multiplier, that enforces the gauge fixing condition, the worldsheet action can be cast into various guises.   
We investigate the renormalization of this theory using a non--linear background/quantum split by employing a normal coordinate expansion adapted to the gauge--fixed theory.
The propagator of the doubled coordinates contains a projection operator encoding that half of them do not propagate. 
We determine the doubled target space equations of motion by requiring one--loop Weyl invariance. 
Some of them are generalizations of the conventional sigma model beta--functions, while others seem to be novel to the doubled theory: 
In particular, a dilaton equation seems related to the strong constraint of double field theory. 
However, the other target space field equations are not identical to those of double field theory.

\newpage 

\setcounter{page}{1}
\tableofcontents
\newpage

\section{Introduction}
\label{sc:intro} 

%
String theory offers a different perspective on the geometry of the target space than an ordinary quantum field theory does. The reason is that strings can wind around compact dimensions and thus probe the target space with both Kaluza--Klein and winding modes whereas point particles only have the former at their disposal. Therefore, strings can be sensitive to so--called non--geometric backgrounds  \cite{hmw02,dh02,fww04,stw05}, which do not admit any standard geometric interpretation and could never be detected by point particles alone. 

%
In particular, this stringy perspective on geometry is reflected by the existence of T--duality in string theory~\cite{Buscher.1987,Buscher.1988} which relates string theories on different compact backgrounds to each other. In the simplest eponymous case, T--duality relates strings on a circle of radius $R$ with one on a circle of radius $\ga'/R$ by exchanging Kaluza--Klein and winding modes. T--duality has been generalized to more general backgrounds and superstring theory, see e.g.~\cite{Bergshoeff.1995}. Because of T--duality's continued success in relating different string theories, there has been a large effort to make this duality a manifest symmetry both on the level of the effective field theory as well as on the worldsheet.

%
On the level of the effective theory this has been accomplished by the framework of double field theory \cite{s93,hz09,hhz10a,hhz10b}, for a recent review see~\cite{Aldazabal.2013}, in which the standard target space coordinates are accompanied by dual coordinates whose Kaluza--Klein modes represent the winding modes of string theory. In this process the number of target space dimensions is doubled unless a suitable constraint is imposed. Usually, one imposes the so--called strong constraint which requires all objects of the theory (and products thereof) to be elements of the kernel of a particular non--linear differential operator.

%
On the level of the worldsheet theory there have been various attempts 
to make T-duality manifest. (For an overview see e.g.\ Ref.~\cite{Berman:2013eva}.) The earliest ones go back to Tseytlin~\cite{t90,t91} based on Ref.~\cite{Floreanini:1987as,Duff:1989tf}, Siegel~\cite{s93} and Hull~\cite{h04}. In these worldsheet theories the number of coordinate fields has been doubled. (Also conjugate momenta have been included in an associated membrane action to describe non--geometry backgrounds on the worldsheet, see e.g.~\cite{Halmagyi:2009te,Halmagyi:2008dr}.) In particular, this has led to so--called T--folds~\cite{h06,hr09}. In Hull's approach~\cite{h06,Hull.2005} a constraint to half the number of degrees of freedom is implemented by hand. In Tseytlin's construction~\cite{t90,t91,dp08} the reduction is implemented by interpreting the coordinate fields and their duals as each others conjugate momenta. Unfortunately, in the course of Tseytlin's construction manifest Lorentz invariance is lost unless additional constraints are enforced. Typically these constraints are stronger than those necessary to ensure on--shell Lorentz invariance~\cite{sst09} and are motivated to get the number of degrees of freedom correct~\cite{Pasti:1996vs}. This complicates the derivation of the corresponding target space equations of motion, see e.g.~\cite{bct07,bt07,adp09,sst09} and~\cite{c11}. Another interesting approach can be found in \cite{hsz13}.

%
Recently, two of us suggested a sigma model for a doubled geometry\footnote{Doubled geometry here means a $2D$-dimensional manifold equipped with a metric and an anti--symmetric field.
}~\cite{Nibbelink:2012} that incorporates the necessary constraints on the level of the worldsheet but contrary to Tseytlin's approach is nevertheless manifestly Lorentz invariant. This theory was motivated as follows: As was observed in~\cite{rt98} for constant metric and Kalb--Ramond backgrounds Buscher's Lorentz invariant gauge theory~\cite{Buscher.1987,Buscher.1988} leads to Tseytlin's description by employing a non--Lorentz invariant gauge fixing. However, it is also possible to implement a Lorentz invariant one. This constraint in principle leads to problematic chiral bosons, see e.g.\ Refs.~\cite{Srivastava:1989zp,Harada:1990fn}, however, we argued that these are canceled by the ghost fields in a proper BRST quantization of the theory.  When the Lorentz invariant gauge fixing is implemented, the remaining gauge field component only appears linearly in the action, which allows to reinterpret this gauge field component as a Lagrange multiplier which itself fixes a gauge symmetry. This gauge symmetry shows that half of the doubled coordinates are redundant and remain present even for non--constant metric and Kalb--Ramond backgrounds as well. In~\autoref{tb:DoubledObjects} we give an overview of the various objects that play important roles within this doubled worldsheet theory and briefly describe both their worldsheet and target space interpretation. 

%
In this paper we investigate the renormalization of this theory exploiting similar methods that were used for the standard sigma model, see e.g.\ Refs.~\cite{Callan:1985ia,Hull:1985rc}. To this end we describe how the suggested gauge fixing procedure is implemented at the quantum level via the BRST quantization procedure involving Faddeev--Popov ghosts. Furthermore, we show that it reduces the theory to the correct number of degrees of freedom both on the worldsheet and in the target space. Furthermore, we show how it is possible to rewrite the theory in different guises that implement $O(D,D)$ transformations either invariantly or covariantly. T--duality then appears as manifest feature of the doubled geometry within this worldsheet theory. However, the field equations we derive are not identical to those of double field theory. There is a good reason for this: By construction, our doubled theory realizes the doubling off-shell on the worldsheet, as a consequence it is invariant under $2D$ dimensional diffeomorphisms. This is not the same gauge symmetry as is realized in double field theory.

\begin{table}
{\renewcommand{\arraystretch}{1.5}
\begin{center}
\begin{tabular}{|cl||cc|p{5.7cm}p{5.7cm}|}
\hline 
\multicolumn{2}{|c||}{\textbf{Object}} & \multicolumn{2}{|c|}{\textbf{\#(comp.)}} &\textbf{Worldsheet interpretation} & \textbf{Target space interpretation} 
\\\hline\hline 
$Y$ & 
& $2D$& &   Doubled coordinate fields 
& Doubled geometry coordinates 
\\ \hline 
& $X$ 
& & $D$ & Worldsheet coordinate fields 
& Target space manifold coordinates 
\\
&$\cK$ 
& & $D$ & Gauge transformations 
& Doubled geometry Killing vectors and projectors  
\\ \hline \hline 
$\cE$ & 
& $4D^2$ & & Kinetic and WZ terms of the 
& Doubled geometry metric $\cG$ and 
\\[-1ex]  
& & & & doubled worldsheet theory  & antisymmetric tensor field $\cC$ 
\\ \hline 
& $\tilde \cE \cong E$ 
& &$D^2$ & Projected version of the doubled  
& Target space metric $g$ and 
\\[-1ex] 
& & & & kinetic and WZ terms & antisymmetric tensor field $b$
\\
& $\cZ$ 
& & $D^2$ & Gauge fixing parameters & \multirow{3}{*}{
$\left\}\vphantom{\rule{0pt}{1.3cm}}\right.$ 
Non--physical parts of $\mathcal{E}$} 
\\[-1ex] 
& & & &  (modulo $\rho$--trans\-formations) & 
\\ 
& $\cQ$ 
& & $2D^2$ & BRST ghost transformation
& 
\\[-1ex]  
& & & &  parameters  & 
\\ \hline 
\end{tabular}
\end{center}}
\caption{ \label{tb:DoubledObjects} 
The central objects of the doubled worldsheet theory with a brief description of both their worldsheet and target space interpretation are collected in this Table. In addition it gives the number of components of these objects. }
\end{table}

\subsection*{Paper summary} 

In Section~\ref{sc:doubleWS} we describe the Lorentz invariant doubled worldsheet theory introduced in Ref.~\cite{Nibbelink:2012}. We review the construction of the model identifying the gauge symmetry which reduces the number of degrees of freedom. The symmetries that this model exhibits are discussed. In addition to multiplicative redefinitions of the Lagrange multiplier, that implements the gauge fixing, we focus on the special transformation that allows to cast the model into different forms. In the final Subsection~\ref{sc:BRST} we investigate the general consequences of enforcing BRST invariance. In particular, we show that the gauge symmetries are encoded by Killing vectors which have particular projection properties.  In the final Subsection~\ref{sc:Stdouble} we describe some special forms of the worldsheet theory making various symmetries manifest. 

Section~\ref{sc:covFey} develops the covariant Feynman rules for the Lorentz invariant doubled worldsheet theory. The background field method applied to non--linear $\sigma$ models and the normal coordinate expansion are adapted to our doubled theory. Covariant derivatives of the gauge fixing Lagrange multiplier which are needed for the covariant expansion are determined by requiring covariance w.r.t.\ its multiplicative redefinitions. In Subsection~\ref{sc:Props} the propagators of the doubled theory are determined. 

Section~\ref{sc:oneloop} is devoted to the study of the renormalization of the Lorentz invariant worldsheet with doubled coordinate fields and the derivation of the resulting target space dynamics. To this end we determine the divergent contributions to the effective action using the Feynman rules derived in the previous Section. By demanding Weyl invariance on the quantum level in Subsection~\ref{sc:TSeom} we determine the target space equations of motion for the Lorentz invariant doubled worldsheet theory.
  
In Section~\ref{sc:TSdiscussion} we discuss the target space interpretation of our worldsheet theory with doubled coordinate fields both at the classical and the quantum level. Section~\ref{sc:outlook} concludes this paper with an outlook on open questions. 
 
In Appendix~\ref{sc:covExp} we have collected some details of the covariant expansion employed in Section~\ref{sc:covFey}. Appendix~\ref{sc:DimReg} contains brief computations of the relevant divergent one--loop integrals within dimensional regularization.

\subsection*{Acknowledgements}

We would like to thank 
Andreas Deser, 
Falk Hassler,
Olaf Hohm 
and Dieter L\"ust 
for very helpful discussions. 
This work has been supported by the LMUExcellent Programme.

 \section{Worldsheet description with doubled target space coordinates}
 \label{sc:doubleWS} 
 
In this Section we introduce the doubled worldsheet theory that is under investigation in this paper. The main players of this theory have been collected in~\autoref{tb:DoubledObjects}. To facilitate the comparison with the standard sigma model description, we recall its basic properties first.

\subsection{Standard sigma model}


The standard sigma model for coordinate fields $X^\gm$, $\gm = 0, \ldots D-1$ of the bosonic string is given by 
\equ{ \label{standardsigmaact} 
S = \int \d^2\sigma\,  \der_L X^T E\, \der_R X 
~, 
}
where $\der_a = \der_{L/R} = (\der_0 \pm \der_1)/\sqrt 2$ denote derivatives w.r.t.\  the light--cone coordinates, $\gs _a = \gs_{L/R} = (\gs_0\pm\gs_1)/\sqrt2$, on the string worldsheet.  In addition, $E_{\gm\gn}(X) = g_{\gm\gn}(X) + b_{\gm\gn}(X)$, where $g_{\gm\gn}$ and $b_{\gm\gn}$ represent the metric on a $D$--dimensional target space $M$ and anti--symmetric Kalb--Ramond field with field strength $H_{\gm\gn\gk} = 3\, \der_{[\gm} b_{\gn\gk]}$. \footnote{The symmetrization and anti--symmetrization of indices denoted by $(\gm_1\ldots\gm_n)$ and $[\gm_1\ldots\gm_n]$, respectively, include a symmetrization factor $1/n!$.}


This action is invariant under conformal transformations which in this representation take the form
\equ{ \label{conftrans} 
\gs_L \ra \gs_L' = h_L(\gs_L)~, 
\qquad 
\gs_R \ra \gs_R' = h_R(\gs_R)~, 
}
where $h_L$ and $h_R$ are two in principle independent holomorphic functions of $\gs_L$ and $\gs_R$, respectively. Field redefinitions of the coordinate fields induce diffeomorphism transformations of the metric and $b$--field: 
\equ{ 
X^\gm \ra f^\gm(X)~, 
\qquad 
g \ra (\der f)^T \, g\, (\der f)~, 
\quad 
b \ra (\der f)^T \, b\, (\der f)~, 
}
for general functions $f^\gm(X)$ and $(\der f)^\gm{}_\gn = \der_\gn f^\gm$. 

\subsection{Doubled worldsheet theories}
\label{sc:doubledWS} 

In \cite{Nibbelink:2012} it was shown that the standard sigma model can be related to a theory with twice the number of coordinates which we refer to as $Y^m$ with $m=0,\ldots, 2D-1$. Given that half of the coordinates are redundant, a gauge transformation was proposed there. The most general form of this gauge transformation is given by 
\equ{ \label{gaugesym}
\gd_\gx Y = \cK(Y)\, \gx~, 
}
where $\gx_{\ga}(\gs)$ are $D$ independent local, i.e.\ worldsheet coordinate dependent, gauge parameters labeled by $\ga=0,\ldots D-1$. Since the composition of two such gauge transformations should itself be a gauge transformation, the so--called Killing vectors $\cK^\ga(Y)$ need to satisfy 
\equ{ \label{KillingAlgebra} 
\cK^{m\alpha}{}_{,p}\, \cK^{p\beta} - \cK^{m\beta}{}_{,p}\, \cK^{p\alpha} = f^{\alpha\beta}{}_\gamma(Y)\, \cK^{m\gamma} \ .
}
The structure coefficients $f^{\alpha\beta}{}_\gamma(Y)$ of their algebra may in general vary over the doubled manifold $\cM$.


In view of this gauge symmetry \eqref{gaugesym} the most general action for the doubled coordinates has to involve some gauge fixing 
\equ{ \label{eq:act} 
S  = \int \dsigma \Big( \frac 12\, \dl Y^T \,\cE\,\dr Y + \cW_L\, V_R \Big) 
~, 
}
with $\mathcal{E}_{mn} = \G_{m n} +\mathcal{C}_{m n}$. The matrix function $\G_{m n}$ can be thought of as the metric on a $2D$--dimensional manifold $\cM$. $\cH_{pmn} = 3\, \der_{[p}\mathcal{C}_{mn]}$ denotes the field strength of the anti--symmetric tensor field $\mathcal{C}_{m n}$. To define a proper quantum theory the $\mathcal{C}$ term is subject to a quantization condition.
In addition, $V_R^\gm$, $\gm=0,\ldots, D-1$, act as the Lagrange multiplier fields, since their equations of motion require that 
\equ{  \label{GFfunct} 
\cW_L = \dl Y^T\, \cZ(Y)~,  
}
is set to zero identically, thereby classically enforcing the gauge fixing. These equations fix all gauge invariances provided that the $2D\times D$ matrix, $\cZ_{m\gm}$,  is chosen such that
\equ{ \label{fullyFixed}
N = \cK^T \cZ~ 
}
is an invertible $D\times D$--matrix. In principle, one could consider more involved gauge fixing conditions, but this is the most general choice that is compatible with the conformal symmetries \eqref{conftrans}. As observed in \cite{Nibbelink:2012} this is not a complete gauge fixing, hence the corresponding ghost sector cannot be ignored (and is discussed below in subsection \ref{sc:BRST}.)

The representation of this doubled worldsheet action is far from being unique because one can perform various redefinitions of the fields on the worldsheet, namely: 
\enums{
\item {\bf Doubled diffeomorphisms}:  
\\[1ex]  
On the doubled coordinates $Y^m$ we can perform generic field redefinitions $Y \ra \cF(Y)$ of the doubled coordinates  which induce doubled diffeomorphisms (with $(\der \cF)^m{}_n = \der_n \cF^m$):
\equ{ \label{doubleddiffs}
\cG \ra (\der \cF)^{-T}\, \cG\, (\der \cF)^{-1}~,
\quad 
\cC \ra (\der \cF)^{-T}\, \cC\, (\der \cF)^{-1}~, 
\quad 
\cZ \ra (\der \cF)^{-T}\, \cZ~, 
\quad 
\cK \ra (\der \cF)\, \cK~, 
}
\item {\bf Redefinitions of the Lagrange multiplier}:  
\\[1ex]  
The Lagrange multiplier $V_R$ can be redefined by arbitrary matrix multiplications:   
\equ{ \label{rhotrans}
V_R \ra \gr(Y)\, V_R~:
\quad 
\cZ \ra \cZ\, \gr^{-1}~, 
}
where $\gr^\mu{}_\nu(Y)$ is a $D\times D$--matrix function of $Y$. 
\item {\bf Shifts of the Lagrange multiplier}: 
\\[1ex]  
The conformal transformations also allow that the Lagrange multiplier $V_R$ is shifted as: 
\equ{ \label{Utrans} 
V_R \ra  V_R + \cU(Y)\, \dr Y~:
\quad 
\cE \ra  \cE - 2\, \cZ\, \cU~, 
}
where $\cU^\gm{}_m(Y)$ is a $D\times 2D$--matrix function of $Y$. 
\item {\bf Redefinitions of the Killing vectors}: 
 \\[1ex]  
 We can allow for transformations $\gx \ra \go(Y)\, \gx$ that redefine the gauge parameters, consequently: 
\equ{ \label{omegatrans}
\cK \ra \cK\, \go^{-1}~, 
\quad 
f^{\ga\gb}{}_\gg \ra 
(\go){}_\gg{}^\gn\, 
f^{\gk\gl}{}_\gn\, 
(\go^{-1}){}_\gk{}^\ga\, 
(\go^{-1}){}_\gl{}^\gb
+  
(\go){}_\gg{}^\gn \, 
(\go^{-1}){}_\gn{}^{[\ga}{}_{,p}\, (\go^{-1}){}_\ge{}^{\gb]}\,
\cK^{p\ge}~, 
}
where $\go_\ga{}^\gb(Y)$ is a $D\times D$--matrix function of $Y$. 
}

\subsection{General BRST transformation}
\label{sc:BRST} 


The transformations identified above all stem from possible redefinitions of the field variables and the gauge parameters. The gauge transformation \eqref{gaugesym} was not included, because by means of the gauge fixing term it is not a symmetry of the action \eqref{eq:act} anymore. More importantly, depending on the detailed form of the matrix $\cE$, the kinetic terms in \eqref{eq:act} are not even gauge invariant by themselves, but only invariant upon using the gauge fixing constraint. 


In a full--fledged off--shell quantum description of the gauge symmetries within the path integral formalism after Faddeev--Popov (FP) gauge fixing, these are reincarnated as BRST transformations $\gd_\ge$: The gauge parameters $\gx_\ga$ are replaced by $\ge\, c_\ga$ where $\ge$ is a constant fermionic variable and $c_\ga$ ghost fields. In addition, to each gauge fixing condition $\cW_{L\ga}$ a $b_R^\ga$ ghost is associated. The various fields and parameters in the FP gauge fixed theory can be classified according to their ghost charge $Q$ and their right--moving conformal weight $R$: 
\equ{ \label{BRSTcharges} 
\begin{array}{|r||c|c|c|c|c|c|c|}
\hline 
\text{Field}  & Y & V_R & c & b_R & \partial_R Y & \partial_R c & \varepsilon \\\hline\hline 
Q & 0 & 0   & 1 & -1  & 0            & 1            & -1          \\\hline
R & 0 & 1   & 0 & 1   & 1            & 1            & 0 \\\hline 
\end{array} 
}
There are two fundamental properties of the BRST transformations: i) they are nilpotent and ii) they leave the full quantum action, including the ghost term, invariant: 
\equ{ \label{BRSTact} 
S = \int \d^2\sigma \Big( \frac{1}{2}\, \partial_L Y^T \cE\, \partial_R Y + \cW_{L}\, V_R + \delta_c \cW_{L}\, b_R \Big)~,
}
where $\cW_L$ is given in \eqref{GFfunct}. 


In the following we analyze the structure of the most general BRST transformations compatible with the ghost charges and conformal weights as given in the table~\eqref{BRSTcharges}. The most general transformation of the coordinates $Y^m$ reads 
\equ{
\delta_\varepsilon Y^m = \varepsilon\, \cK^{m\ga}(Y) c_\ga \ .
}
which is precisely the classical gauge transformation \eqref{gaugesym} with the gauge parameters replaced by $\gve$ times the $c$--ghost. The nilpotency of $\delta_\varepsilon$ applied onto $Y$ fixes how the ghost $c_\ga$ transforms: 
\equ{
\delta_{\varepsilon'} \delta_\varepsilon Y^m \overset{!}{=} 0 
\qquad \Rightarrow \qquad 
\delta_\varepsilon c_\gamma = \frac{1}{2}\, \varepsilon\, f^{\alpha\beta}{}_\gamma\, c_\alpha c_\beta \ ,
}
where we used the algebra of the Killing vectors \eqref{KillingAlgebra}. 
Using the nilpotency on $c$ determines an extra condition:
\equ{ \label{genjacobi}
f^{\kappa[\alpha}{}_\lambda f^{\beta\gamma]}{}_\kappa + \cK^{p[\alpha} f^{\beta\gamma]}{}_{\lambda,p} = 0 \ ,
}
which is precisely the Jacobi identity for non--constant structure functions $f^{\alpha\beta}{}_\gamma$. In particular, \eqref{genjacobi} reduces to the known Jacobi identity for constant structure coefficients. The reason why we allow for non--constant structure coefficients in the first place is that it is possible to perform local redefinitions of the gauge parameters \eqref{omegatrans}, and consequently of the $c$--ghosts as $c \ra \go(Y) c$, which would turn constant structure coefficients into field dependent ones. 


To determine the transformation rule of the ghost field $b_R^\mu$ we consider the most general ansatz
\equ{
\delta_\varepsilon b_R^\mu = \varepsilon\, A^\mu{}_\nu(Y)\, V_R^\nu + \varepsilon\, B^{\mu\beta}{}_\nu(Y)\,  c_\beta b_R^\nu + \varepsilon\, \cQ^\mu{}_m(Y)\, \partial_R Y^m \ ,
}
which contains all possible terms with a ghost charge $Q = -1$ and a Lorentz charge $R = 1$ according to the table given in equation \eqref{BRSTcharges}. We can make two simplifications: 1) The matrix function $A^\mu{}_\nu$ can be absorbed in the definition of $b_R^\nu$. 2) Since the second term only involves the ghost fields, it can never affect the structure of the kinetic terms of the coordinate fields, therefore, we can set $B^{\mu\beta}{}_\nu = 0$ without restricting the double target space properties encoded in this theory. The BRST transformation of the $b_R$--ghost field then reduces to 
\equ{ \label{BRST_bR} 
\delta_\varepsilon b_R^\alpha = \varepsilon \,V_R^\alpha + \varepsilon\,  \cQ^\alpha{}_m\, \partial_R Y^m \ .
}
Finally, from $\delta_{\varepsilon'}\delta_\varepsilon b_R^\alpha \stackrel{!}{=} 0$ we read off the general transformation behavior of $V_R^\alpha$:
\equ{ \label{BRST_VR} 
\delta_\varepsilon V_R^\alpha = 
- \varepsilon \big( \cQ^{\alpha}{}_{m,p} \cK^{p\beta} + \cQ^\alpha{}_p \cK^{p\beta}{}_{,m} \big) c_\beta \partial_R Y^m 
- \varepsilon\, \cQ^\alpha{}_p \cK^{p\beta}\,  \partial_R c_\beta \ .
}
This transformation of $V_R$ appears to be much more complex than the standard BRST--transformation of a Lagrange multiplier field enforcing the gauge fixing conditions, which simply reads $\gd_\gve \tV_R = 0$. However, it is easy to confirm that by the  $\cU$--transformation~\eqref{Utrans}, setting $\cU = \cQ$, we precisely obtain the Lagrange multiplier field $\tV_R$ which is BRST inert. 


Having determined the complete set of nilpotent BRST transformations, we are now in the position to explore the consequences of the requirement of BRST invariance of the quantum action~\eqref{BRSTact}. Using the above relations, we find
\equa{ 
\delta_\varepsilon S = \varepsilon \int \d^2\sigma \Big\{ & 
\partial_L Y^m \partial_R c_\beta\, \cK^{p\beta} \, 
\Big( \frac{1}{2} \cE_{mp} - \cZ_{m\mu} \cQ^\mu{}_p \Big) 
+ \partial_L c_\beta \partial_R Y^m\, \cK^{p\beta} \, 
\Big( \frac{1}{2} \cE_{pm} - \cZ_{p\mu} \cQ^{\mu}{}_m \Big)  
\non \\[1ex]
& + \partial_L Y^m \partial_R Y^n c_\beta \, 
\Big( \frac{1}{2} \cK^{p\beta}{}_{,m} \cE_{pn} + \frac{1}{2} \cK^{p\beta}{}_{,n} \cE_{mp} + \frac{1}{2} \cK^{p\beta} \cE_{mn,p}  
\\[1ex] 
& - \cZ_{m\mu} \cQ^\mu{}_{n,p} \cK^{p\beta} 
- \cZ_{m\mu,p} \cQ^\mu{}_n \cR^{p\beta} 
- \cZ_{m\mu} \cQ^\mu{}_p \cK^{p\beta}{}_{,n} 
- \cZ_{p\mu} \cQ^\mu{}_n \cK^{p\beta}{}_{,m} \Big) \Big\} \ . \non 
}
The conditions for the BRST invariance of the quantum action can be cast in the following simple form 
\begin{subequations} \label{BRSTconditions}
\equ{ 
\cK^{p\alpha}{}_{,m}\, {\tilde\cE}_{pn} 
+ \cK^{p\alpha}{}_{,n}\, {\tilde\cE}_{mp} 
+ \cK^{p\alpha}\, {\tilde\cE}_{mn,p} = 0~, 
\label{KillingEqs}
\\[2ex] 
{\tilde\cE}\, \cK = \cK^{T} {\tilde\cE} = 0~. 
\label{ProjectionEqs} 
}
\end{subequations} 
Since the former is of the form of standard Killing equations while the latter can be viewed as projection equations we will often refer to them collectively as the {\em projective Killing equations}. Here we have introduced the matrix $\tilde\cE$, defined by 
\equ{ \label{ExpcE} 
\qquad 
\cE = \tilde\cE + 2\, \cZ\, \cQ~.
}
Notice that the relation between $\tilde\cE$ and $\cE$ is precisely a $\cU$--transformation \eqref{Utrans} with $\cU=\cQ$.  Hence, precisely when $\cE$ takes its simplest form, the BRST--transformation of $V_R$ is trivial. 




The first equation in \eqref{BRSTconditions} is the Killing equation for the vectors $\cK^\ga$ w.r.t.\ ${\tilde\cE}$. This justifies calling the $\cK^\ga$ Killing vectors. The remaining two equations in \eqref{BRSTconditions} imply that $\cE$ describes the same number of target space degrees of freedom as the matrix $E$ of the standard sigma model \eqref{standardsigmaact}. Indeed,  these equations tell us that ${\tilde\cE}$ is perpendicular to all Killing vectors from both sides. This means that only $D^2$ of the components of the $2D\times 2D$--matrix ${\tilde\cE}$ are independent. Even though this shows that  the matrix $E$ of the standard sigma model and the matrix $\cE$ of the doubled theory have the same number of independent components, in general the relation between these two objects might be very complicated. (A more detailed account on the reduction of the degrees of freedom can be found in \autoref{sc:TSdiscussion}.)


In the light of the projective Killing equations~\eqref{BRSTconditions} it might seem disturbing that the structure coeficients were allowed to be non--constant, because in general the Killing equations~\eqref{KillingEqs} are not satisfied by $f \cK^\ga$ when $f$ is a generic target space function. Indeed, inserting this expression into the equation~\eqref{KillingEqs} one finds that additional terms like $(\partial_m f) \, \cK^{p\gb} {\tilde \cE}_{pn}$ arise, because the derivative may also hit the function $f$. However, because of the additional perpendicularity conditions~\eqref{ProjectionEqs}, these terms vanish. 
Thus, contrary to the generic case, one can here allow for non--constant coefficients in the algebra of Killing vectors. This ensures that the theory is compatible with the transformation~\eqref{omegatrans} in which the $\cK^\ga$ may turn into non--constant linear combinations of the old Killing vectors.

\subsection{Special forms of the doubled worldsheet theory}
\label{sc:Stdouble} 

The formalism developed so far takes the idea of a doubled worldsheet to the extreme in the sense that invariance under $2D$--dimensional diffeomorphisms \eqref{doubleddiffs} and $\rho$-transformations \eqref{rhotrans} is manifest. However, to see the physical content more clearly, it is useful to choose particular representations of the theory. In this section we discuss some of such forms that make either $O(D,D)$ symmetry or $D$--dimensional diffeomorphisms manifest.

\subsubsection*{Manifest global $\mathbf{O(D,D)}$ cov/invariance}


As the Killing vectors $\cK^\ga$ are associated to the gauge transformations that leave the doubled worldsheet theory inert, they locally point into the $D$ redundant directions. Hence by a change of doubled coordinates one can ensure that these directions correspond to the dual coordinates. This is possible because the algebra~\eqref{KillingAlgebra} of the Killing vectors $\cK^\ga$ closes, so that they span a so--called involutive distribution. Then by Frobenius' theorem \cite{Warner:1983}, around every point one can find a coordinate chart such that, locally, $\cK$ is of the form
\equ{ \label{Killingstandard} 
\cK = \left(\begin{array}{c} 0 \\ K \end{array}\right)~;
\quad\text{and set}\quad  
\cZ = \left(\begin{array}{c} E \\ \Id_D \end{array}\right)~
}
where $K$ is a $D \times D$ matrix function. Moreover, the fact that the Killing vectors are linearly independent at every point ensures that $K$ is invertible and thus the consistency condition \eqref{fullyFixed} is satisfied. 

Now redefining the Killing vectors as in \eqref{omegatrans}, with the special choice $\go = K$, we can even bring $\cK$ into the simple form
\equ{ \label{Killingstandard2}
\cK = \left(\begin{array}{c}0 \\ \Id_D \end{array}\right) \ ,
}
as considered in \cite{Nibbelink:2012}. This form of the Killing vectors identifies the physical coordinates $X^\gm$ of the sigma model with the upper half of the coordinates of $Y^m$, while the lower half is identified with the (redundant) dual coordinates $\tilde X_\gm$. 


Moreover,  in the standard form~ \eqref{Killingstandard2} of the Killing vectors, the matrix $\tilde\cE$ is forced to be of the form 
\equ{ \label{tcEstandard} 
\tilde\cE = \left(\begin{array}{cc} 2\,E & 0 \\0 & 0\end{array}\right)~, 
}
in order to satisfy the projection conditions~\eqref{ProjectionEqs}. 
Using particular $\cU$--transformations~\eqref{Utrans}, we can represent the metric of the doubled worldsheet in various forms as indicated in~\autoref{tb:StForms}. (The anti--symmetric matrices $\cC$ that arise in the $O(D,D)$ in-- and covariant forms in \autoref{tb:StForms} correspond to a mere total derivative on the worldsheet and are therefore non--physical.) This shows in a background independent way that we can locally bring the kinetic terms of the doubled theory into an $O(D,D)$ invariant form, since these arguments did not rely on any specific form of this matrix $E$.

%
\begin{table} 
\equ{ 
\begin{array}{|c||c|c||c|}
\hline 
\textbf{Form}  & \mathbf{\cG} & \mathbf{\cC} & \mathbf{\cU} \\ \hline\hline &&& \\[-2ex] 
\text{Standard sigma model}  & 
\left(\begin{array}{cc} 2\,g & 0 \\0 & 0\end{array}\right) &
\left(\begin{array}{cc} 2\,b & 0 \\0 & 0\end{array}\right) & 
0
\\[-2ex] &&& \\\hline &&& \\[-2ex] 
O(D,D)~\text{invariant} & -\get = -
\left(\begin{array}{cc}0 & \Id_D \\ \Id_D & 0\end{array}\right) & 
\left(\begin{array}{cc}0 & \Id_D \\ -\Id_D & 0\end{array}\right) & 
\left(\begin{array}{cc}\Id & 0\end{array}\right)
\\[-2ex] &&& \\\hline &&& \\[-2ex] 
O(D,D)~\text{covariant}& \cH = 
\left(\begin{array}{cc} 
g-bg^{-1}b & bg^{-1} \\ 
-g^{-1}b & g^{-1} 
\end{array}\right) & 
\left(\begin{array}{cc}0 & \Id_D \\ - \Id_D & 0\end{array}\right)  & 
\frac 12 \left(\begin{array}{cc}\Id_D + g^{-1} b & - g^{-1} \end{array}\right)
\\[-2ex] &&& \\\hline 
\end{array}
\non 
}
\caption{Three different standard forms for the doubled worldsheet theory are indicated and the required $\cU$--transformation \eqref{Utrans} to reach that realization from the form \eqref{tcEstandard}. 
\label{tb:StForms}}
\end{table} 


In the $O(D,D)$ invariant form and for constant backgrounds, the kinetic term of the doubled coordinate fields $Y$ is invariant under global $\cM \in O(D,D)$ transformations 
\equ{
Y \ra \cM^{-T}\, Y~, 
\qquad 
\cM \, \get \,\cM^T = \get~, 
}
while the so--called generalized metric $\cH$ transforms covariantly, i.e.\ 
$\cH \ra \cM \,\cH\, \cM^T$. If one insists on preserving the standard form~\eqref{Killingstandard} for the constraint matrix $\cZ$, one needs to perform a compensating $\rho$--transformation~\eqref{rhotrans} with $\gr = \gg\, E+\gd$.\footnote{This form of $\rho$--matrix is quite reminiscent of the anchor map discussed in e.g.~\cite{Blumenhagen:2012nt,Blumenhagen:2013aia}.} Consequently, the matrix $E$ transforms as 
\equ{\label{nonlinodd}
E \ra (\ga\, E + \gb) (\gg\, E +\gd)^{-1}~, 
\qquad 
\cM = \left(\begin{array}{cc}\ga & \gb \\ \gg & \gd \end{array}\right)~. 
}
%


Even though this formulation makes the global $O(D,D)$ transformations manifest, it is not covariant w.r.t.\ $2D$--diffeomorphisms. This means that in general the specific form \eqref{Killingstandard} only holds within one particular coordinate patch at best. In particular, the renormalization of the constraint does not preserve this choice. This we have verified by applying the one--loop renormalization formulae to be derived in the next sections.

As a side remark, the following should be noted: It is possible to encode $H$-flux in the generalized antisymmetric field $\cC$ such that consequent $O(D,D)$ transformations reveal the whole chain of fluxes \cite{stw05},
\begin{equation}
H_{abc}\ \rightarrow\ f^{a}{}_{bc}\ \rightarrow\ Q_c{}^{ab}\ \rightarrow\ R^{abc} \ .
\end{equation}
This has been worked out in \cite{Nibbelink:2012}, where also the special case of a three-torus with $H$-flux \cite{allp12,allp11} is discussed. Non-trivial monodromies, that turn the backgrounds with $Q$- or $R$-flux into non-geometric ones, appear precisely through the non-linear transformation \eqref{nonlinodd}.

\subsubsection*{Embeddings of $\mathbf{D}$--dimensional diffeomorphisms in doubled diffeomorphisms}


If one comes from or wants to compare with a standard sigma model description, only the $D$--dimensional diffeomorphisms of the coordinates,  
\equ{ \label{Ddiffs} 
X^\mu \ra f^\mu(X)~, 
}
need to be realized explicitly. In principle the $D$--dimensional diffeomorphisms form a subgroup of the $2D$--diffeomorphisms, since we can simply write 
\equ{ \label{StEmDdiffs} 
\cF(Y) = \pmtrx{ F(Y) \\ \wt{F}(Y) }~, 
\quad\text{with}\quad  
F(Y) = f(X)~;~ \wt{F}(Y)=\wt{X}~, 
\quad\text{so that}\quad  
\der \cF = 
\pmtrx{ \der f & 0 \\ 0 &  \Id}~. 
}
However, this does not lead to the expected transformation of the generalized metric
\equ{ \label{Htrafo} 
\cH \ra 
\begin{pmatrix}
\partial f & 0 \\
0 & (\partial f)^{-1}
\end{pmatrix}^T \,
\cH \,
\begin{pmatrix}
\partial f & 0 \\
0 & (\partial f)^{-1}
\end{pmatrix}~. 
}
This form is expected because the generalized metric is only defined in terms of covariant tensors $g, b$ and $g^{-1}$ under $D$--dimensional diffeomorphisms. But such a transformation cannot be obtained from the transformation~\eqref{StEmDdiffs} because the dual coordinates and therefore the dual indices do not transform under it.


One might try to modify the embedding of the $D$--dimensional diffeomorphisms in their 2D analogues to recover the transformation property \eqref{Htrafo} for the generalized metric $\cH$, but it turns out that this is impossible. To see that, let us consider a general ansatz in~\eqref{StEmDdiffs} and enforce the required form of $\der\cF$: 
\equ{ 
\begin{pmatrix}
\partial_X F & \partial_{\wt{X}} F \\
\partial_X \wt{F} & \partial_{\wt{X}} \wt{F}
\end{pmatrix}
\stackrel{!}{=} 
\begin{pmatrix}
\partial f & 0 \\
0 & {\partial f}^{-1}
\end{pmatrix}~.
}
Integrating the off--diagonal equations gives $F(Y)=F(X)$ and
$\wt{F}(Y)=\wt{F}(\wt{X})$. Plugging these results back into the diagonal equations leads to  
\equ{
\partial_{\wt{X}} \wt{F}(\wt{X}) 
= (\partial f)^{-1} (X) 
= 
\big(  \partial_X F(X) \big)^{-1}
~.
}
Since the left--hand--side is a function of $\wt{X}$, while the right--hand--side depends on $X$, this can only be solved for constant 
$\partial_{\wt{X}} \wt{F} =(\partial_X F)^{-1}$ matrices.

The same issue can also be seen directly within the worldsheet theory: If one fixes a $\rho$--gauge such that the constraint matrix $\cZ$ takes the form \eqref{Killingstandard}, then this gauge is only preserved if $\tilde X_\mu$ and $V^\mu_R$ redefinitions are correlated. If one in addition wants to require that $E$ transforms as a rank--two tensor under $D$--dimensional diffeomorphisms, then the $\rho$--transformation needs to take the form $\rho^\mu{}_\nu = f^\mu{}_{,\nu}$. This in turn requires that $\tilde X_\mu$ transforms contravariantly, but that cannot be embedded in a $2D$--dimensional diffeomorphism.

\subsubsection*{Manifest $\mathbf{D}$--dimensional diffeomorphism invariance}

However, it is possible to rewrite the theory in such a way that $D$--dimensional diffeomorphisms \eqref{Ddiffs} appear to be manifestly realized, which are unrelated to $2D$--diffeomorphisms. To this end, we require that the dual coordinate $\tilde X_\mu$ and the Lagrange multiplier $V_R^\mu$ transform contra-- and covariantly, i.e.: 
\equ{ 
\tilde X_\mu \ra \tilde X_\nu\, (\der f^{-1})^\nu{}_{\mu}~, 
\quad 
V_R^\mu \ra (\der f)^\mu{}_\nu\, V_R^\nu~, 
\qquad 
(\der f)^\mu{}_\nu = f^\mu{}_{,\nu}~, 
}
respectively, and we introduce $D$--dimensional diffeomorphism covariant worldsheet derivatives, 
\equ{ 
D_a Y^m = (\cA\, \der_a Y)^m = 
D_a \left(\begin{array}{c}X^\mu  \\ \tilde X_\nu \end{array}\right)
= 
\left(\begin{array}{cc} \delta^\mu{}_\gk & 0 \\ 
- \gg^\gr_{\gn\gk}\, \tilde X_\gr & \delta_\nu{}^\gl \end{array}\right)
\left(\begin{array}{c} \der_a X^\gk \\ \der_a \tilde X_\gl \end{array}\right)
~, 
}
where $\gg^\gr_{\gl\gk}(X)$ defines the connection in $D$ dimensions, e.g.\ the Levi--Civita connection w.r.t.\ the metric $g_{\gm\gn}(X)$. The doubled worldsheet theory can then be written in terms of $D$--dimensional diffeomorphisms as 
\equ{ \label{eq:actionDdimcov}
S = \int \dsigma\, 
\Big\{
\frac 12\, D_L Y^T\, \cE_D\, D_R Y + 
D_L Y^T\, \cZ_D\, V_R
\Big\}~, 
}
where 
\equ{
\cE = \cA^T\, \cE_D\, \cA~, 
\qquad 
\cZ = \cA^T\, \cZ_D~. 
}
This comment should be rather taken as a side remark; we will not use the above rewriting~\eqref{eq:actionDdimcov} of the theory in the following.

\section{Covariant Feynman rules}
\label{sc:covFey} 
 
In this section we set up the quantization of the doubled worldsheet theory described in the previous section. In particular we are interested in the one--loop renormalization of the kinetic terms of the doubled coordinate fields $Y$ and the constraint term involving $V_R$. To this end, we employ a covariant quantum/background splitting of the coordinate fields $Y$ and the Lagrange multiplier. We do not consider the renormalization of the ghost action as we treat the ghost fields as pure quantum objects. Since we are only interested in the one--loop renormalization, it suffices to only determine the two--point vertices for the quantum fields. 
 
 
 As discussed in Subsection \ref{sc:doubledWS} the worldsheet theory on the doubled manifold $\cM$ possesses various symmetries. In the following we set up a background/quantum splitting that is covariant w.r.t.\ doubled diffeomorphisms \eqref{doubleddiffs} and $\rho$--transformations \eqref{rhotrans}. However, we deliberately do not aim to set up a covariant background/quantum splitting w.r.t.\ the $\cU$--transformation \eqref{Utrans}. 


To employ a covariant formalism w.r.t.\ doubled diffeomorphisms we need to have an invertible metric $\cG$ on the doubled manifold $\cM$. However, as can be seen explicitly in~\eqref{tcEstandard} when the Killing vectors are taken to be as in~\eqref{Killingstandard}, the metric $\cG$ on the doubled space may not be invertible. However, by a suitably chosen $\cU$--transformation given in~\eqref{Utrans} we can turn the non--invertible $\tilde\cG$ into an invertible $\cG$. Which form this metric $\cG$ takes is far from unique, since it strongly depends on which $\cU$ one chooses. It may therefore appear that there are huge ambiguities how quantum corrections manifest themselves.  To summarize we need to use the $\cU$--transformation to ensure that $\cG$ is invertible so as to set up a doubled diffeomorphism covariant formalism.

\subsection{Covariant derivatives}
\label{sc:CovDers} 


For the background covariant formalism we need to introduce various appropriate covariant derivatives. We denote by $\cD_a$ and $\cD_m$ the covariant derivatives w.r.t.\ doubled diffeomorphisms alone on the worldsheet and the doubled target space, respectively. In other words, $\cD_m$ represents the standard Levi--Civita connection on $\cM$. Furthermore, the derivatives $\nabla_a$ and $\nabla_m$ are covariant both w.r.t.\ doubled diffeomorphisms as well as $\rho$--transformations. (For objects that do not transform under the $\rho$--transformations at all, of course these derivatives simply coincide.) Concretely, we have 
\equ{ 
\nabla_a Y^m = \cD_a Y^m = \der_a Y^m~, 
\qquad 
\nabla_b \nabla_a Y^m = \cD_b \cD_a Y^m 
= \big( \delta^m{}_n\, \der_b + \gG^m_{kn}\, \der_b Y^k ) \der_a Y^n~.
}
where $\gG^m_{kn}$ are the Christoffel symbols associated to the metric $\cG$. On doubled target space tensors $T^m, T_{mn}, \ldots$, we similarly have 
\equ{ 
\nabla_p T^m = \cD_p T^m = \der_p T^m + \gG^m_{pk}\, T^k~, 
\qquad 
\nabla_p T_{mn}= \cD_p T_{mn} =\der_p T_{mn} - \gG^k_{pm}\, T_{kn} - \gG^l_{pn}\,T_{ml}~, 
} 
etc. In particular, as the Levi--Civita connection $\cD_m$ is metric compatible, one has
\equ{ \label{MetricComp} 
\cD_p \cG_{mn} = 0~, 
}
and  the classical equation of motion corresponding to \eqref{eq:act} can be cast in the form
%
%
\equ{ \label{eq:classEOM}
\nabla_L \nabla_R Y^n = - \frac{1}{2} \cG^{n m}\, \cH_{m pq}\,  \dl Y^p \dr Y^q + \cG^{n m} \Big( \cD_{[m} \cZ_{p] \nu} \dl Y^p \vr^\nu + \cZ_{m \nu} \dl \vr^\nu \Big)~.
}
%

The curvature of the doubled geometry is measured by the Riemann tensor
\equ{ \label{Curvature} 
[ \cD_m, \cD_n] T^p = \cR^p{}_{qmn}\, T^q~, 
\qquad 
\cR^p{}_{qmn} = 
\der_m \gG^p_{nq} - \der_n \gG^p_{mq} 
+ \gG^r_{nq}\, \gG^p_{mr} - \gG^r_{mq}\, \gG^p_{nr}~; 
}
and the corresponding Ricci tensor is defined in the standard way $\cR_{mn} = \cR^p{}_{mpn}$. 


To define $\rho$--transformation covariant derivatives we first introduce $2D\times 2D$--matrices 
\equ{ \label{projectors} 
\cP_\parallel = \cZ \, (\cZ^T \cG^{-1} \cZ)^{-1}\, \cZ^T \cG^{-1}~, 
\qquad 
\cP_\perp = \Id_{2D} - \cP_\parallel ~, 
}
which have the following properties 
\equ{ \label{projections} 
\cP_A^2 = \cP_A~, 
\qquad 
\cG\, \cP_A^T \, \cG^{-1} = \cP_A~, 
\qquad 
\Tr[ \cP_A] = D~, 
}
for $A=\parallel, \perp$ and 
\equ{ 
\cP_\parallel + \cP_\perp = \Id~, 
\qquad 
\cP_\parallel \, \cP_\perp = \cP_\perp\, \cP_\parallel  = 0~;  
\qquad 
\cP_\parallel\, \cZ = \cZ~, 
\qquad
\cP_\perp\, \cZ = 0~, 
} 
These properties signify that the operators $\cP_\parallel$ and $\cP_\perp$ are Hermitean w.r.t.\ the metric $\cG$ as well as projecting on two complimentary $D$--dimensional subspaces that are locally parallel/perpendicular to the components of the matrix $\cZ$, respectively. For later use we also introduce the notation 
\equ{ \label{Uinvariants}
\arry{lccl}{
\G_\perp \!\!&=&\!\! \cP_\perp \cG \!\!&= \cG - \cZ \left(\cZ^T \G^{-1} \cZ\right)^{-1}\cZ^T~, 
\\[2ex]  
\G^{-1}_\perp  \!\!&=&\!\! \cG^{-1}\, \cP_\perp \!\!&= 
\G^{-1}  - \G^{-1} \cZ \left( \cZ^T \G^{-1} \cZ \right)^{-1} \cZ^T \G^{-1}~, 
}
\qquad 
{\cZ}_\parallel = \G^{-1} \cZ \left( \cZ^T \G^{-1} \cZ \right)^{-1}~, 
}
inspired by the definition of the projection operators $\cP_\parallel$ and $\cP_\perp$ in \eqref{projections}. Implicitly we assume the notation $\G^{-1}_\perp$ to mean that one first computes the inverse of $\cG$ and after that projects with $\cP_\perp$. (The other way around is meaningless since the projector $\cP_\perp$ is not invertible.) These operators satisfy 
\equ{ 
\cZ^T\, \cG_\perp^{-1} = \cG_\perp^{-1} \cZ = 0~, 
\qquad 
\cZ^T\, \cZ_\parallel = \Id_D~, 
\qquad 
\cZ_\parallel\, \cZ^T = \cG^{-1}\, \cP_\parallel\, \cG~. 
}
%


Using the doubled diffeomorphism covariant derivatives, we can construct a derivative, 
\equ{ \label{rhoCovDer} 
\nabla_a V_R = 
\der_a V_R + 
\cZ_\parallel^T  \cD_m \cZ \, \der_a Y^m 
\, V_R~,  
\qquad 
\cD_m \cZ_{n\ga} = \der_m \cZ_{n\ga} - \gG^k_{mn}\, \cZ_{k\ga}~,  
}
which is covariant under $\rho$--transformations \eqref{rhotrans} as well. It might appear that because of the inversion the factors $\cZ^T \cG^{-1}$ simply drop out here. This is not the case since $\cZ$ is a rectangular $2D\times D$--matrix and not a square matrix. Given this definition, we can determine how the fully covariant derivative $\nabla_m$ acts on $\cZ$ itself: Using that $(\cZ V_R)_m$ transforms as a tensor $T_m$, we infer that the derivatives $\nabla_m$ and $\cD_m$ on $\cZ$ are not the same but closely related
\equ{ 
\nabla_m \cZ  = \cP_\perp\, \cD_m \cZ~, 
\qquad 
\cZ_\parallel^T\, \nabla_m \cZ = 0~, 
} 
where the matrices $\cP_\perp$ and $Z_\parallel$ are given in~\eqref{projectors} and~\eqref{Uinvariants}, respectively.  Further fully covariant derivatives can be determined in a similar fashion, e.g.\  
\equ{ 
\nabla_m \nabla_n \cZ = 
\cP_\perp\, \cD_m \cD_n \cZ 
- 2\, \cP_\perp \cD_{(m} \cZ\, \cZ_\parallel^T\, \cD_{n)}\cZ 
- \cG\, \cZ_\parallel\, \cD_m\cZ^T\, \cG^{-1}_\perp\, \cD_n \cZ~,
} 
using the definitions \eqref{Uinvariants}.

\subsection{Covariant background/quantum splitting}
 \label{sc:CovSplitting} 
 
 
Using the above derivatives we can set up a fully covariant background/quantum splitting following \cite{Hull}. To this end, we define $Y(\gs;s)$ depending on the affine parameter $s \in  [0,1]$ to be a finite geodesic curve on $\cM$ with respect to the Levi--Civita connection $\nabla$, i.e.\ 
\equ{ \label{eq:geodEq}
\nabla_s^2\,  Y^m(s)  =  
\Big(   \gd^m{}_l \,\frac {\der}{\der s}
+ \Gamma^{m}_{kl}\big(Y(\gs;s)\big)\, \dot Y^k(\gs;s) \Big)  \dot Y^l(\gs;s) 
= 0~, 
}
where $\dot Y(\gs;s) = \frac{\der}{\der s} Y(\gs;s) = \nabla_s Y(\gs;s)$,  
subject to the initial conditions
\equ{
Y(\gs; 0) = Y(\gs)~, 
\qquad 
\nabla_s Y(\gs;0) = y(\gs)~. 
}
Here we interpret $Y$ as the background field and $y$ as the covariant quantum field, which transforms as $y \ra (\der \cF)\, y$ under doubled diffeomorphisms. 
%
%
Similarly, we can define a covariant background/quantum splitting for the Lagrange multiplier field $V_R$: We require that $V_R(\gs;s)$ satisfies the equation 
\equ{ \label{VRgeodesic}
\nabla_s^2\, V_R(\gs;s) = 0~, 
\qquad 
\nabla_s\, V_R = 
\Big( \frac {\der}{\der s} 
+ 
\cZ_\parallel^T \cD_m \cZ\, \dot Y^m 
\Big) V_R~.   
}
In this covariant derivative $\nabla_s$ we have omitted the $Y(\gs;s)$ dependence in order to keep the notation readable. Again, the background and the covariant quantum fields, $V_R$ and $v_R$ are encoded via the initial conditions 
\equ{
V_R(\gs; 0) = V_R(\gs)~, 
\qquad 
\nabla_s V_R(\gs;0)  = v_R(\gs)~, 
}
respectively.
%
%
In principle the full quantum fields, $Y_\text{full}$ and $V_{R\,\text{full}}$, can be expanded as 
\equ{ \label{eq:exp}
\arry{rcl}{
Y_\text{full}(\gs) = Y(\gs;1) &=& \dsp 
 Y(\gs) +  y(\gs)  + \sum_{n \geq 2} \frac{1}{n!} 
\frac{\der^n Y}{{\der s}^n}(\gs; 0) ~, 
\\[1ex]  
V_{R\,\text{full}}(\gs) = V_R(\gs;1) &= & \dsp 
 V_R(\gs) +  v_R(\gs)  + \sum_{n\geq 2} \frac{1}{n!}
\frac{\der^n V_R}{{\der s}^n}(\gs; 0) ~, 
}
}
in terms of the background and covariant quantum fields only, by putting further covariant derivatives on the equations \eqref{eq:geodEq} and \eqref{VRgeodesic} we can find expressions for the higher $s$--derivatives. 


Next, we want to obtain the expansion of the action \eqref{eq:act} in terms of the the quantum fields $y$ and $v_R$. In principle this could be obtained by inserting the expansions \eqref{eq:exp} into the action. But this is rather cumbersome since one then has to package things in covariant objects by hand. A more comfortable procedure to obtain this expansion has been developed by  Ref.~\cite{Howe.1988}: Apply the same method that was used to obtain the background/quantum splittings of the full quantum fields to the action itself. To this end we first promote the action to be dependent on the affine parameter $s$ as well: 
\equ{
 S(s) = \int \dsigma \Big\{ 
 \frac 12\, \Big(\G_{m n}(s) +\mathcal{C}_{m n}(s) \Big)\, 
  \dl Y^m(s) \dr Y^n(s) 
  + \dl Y^m(s)\,  Z_{m \nu}(s)\,  V_R^\nu(s)
  \Big\}~,  
}
writing $\G_{m n}(s) = \G_{m n}(Y(s))$, etc.  Note that the original action \eqref{eq:act} is recovered for $s=1$. Therefore, the expansion of this action in terms of the covariant quantum fields is obtained by making a Taylor expansion of $S(s)$ in $s$ around zero and subsequently setting $s=1$: 
\equ{
S = S(1) = 
\sum_{n \geq 0} \frac{1}{n!} \frac{d^n S}{{ds}^n}(0)
~.  
}
Since the action is a scalar quantity, the repeated ordinary $s$ differentiations can be replaced by fully covariant derivatives $\nabla_s$ on the $s$--dependent fields $Y(s)$ and $V_R(s)$. The geodesic equations \eqref{eq:geodEq} and \eqref{VRgeodesic} imply that squares of $\nabla_s$ on these vanish; while single derivatives on $Y(s)$ and $V_R(s)$ give the fully covariant quantum fields $y$ and $v_R$ once $s$ is set to zero. Hence, the main advantage of this procedure is that it directly gives an expansion in terms of covariant objects only.


Applying this procedure gives back the original action at zeroth order in terms of the background fields $Y$ and $V_R$. By definition of the 1PI effective action the first order terms can be ignored, hence, the first relevant order is the second. Since we are only interested in one--loop results in this work, this second order is, in fact, all we need. After some calculations, for details see Appendix \ref{sc:covExp}, we obtain 
\equa{
S_2 =  \int \dsigma \Big\{ &
\frac 12\, \G_{kl}\, \nabla_L y^k \nabla_R y^l  
+ \frac 12\,  \cZ_{m \mu}\,
\Big(  \nabla_L y^m  v_R^\mu - y^m \nabla_L v_R^\mu \Big) 
\non \\[1ex]  
&
+ \frac 12\, \Big[ 
\Big( \cR_{m kl n}  +  \frac 12\, \nabla_{(k} \cH_{l) mn} 
\Big)\, \dl Y^m \dr Y^n 
+ 
\Big( 
\nabla_{(k} \nabla_{l)} \cZ_{m \mu} +  \cR^p{}_{klm}  \cZ_{p \mu}  
\Big) \, 
\dl Y^m V_R^\mu 
\Big] \, y^k y^l 
\non \\[1ex]  
& + \frac{1}{4}\, \cH_{klm}\, 
\Big( \partial_R Y^m \, y^k \nabla_L  y^l - \partial_L Y^m \, y^k \nabla_R  y^l \Big) 
+ \nabla_k \cZ_{l\mu}\, V_R^\mu\, y^k \nabla_L y^l
\non \\[1ex]  
&
+ \Big( 
\nabla_k \cZ_{m \mu} - \frac 12\,  \nabla_m \cZ_{k \mu} 
\Big)\, \der_L Y^m \,   
y^k  v_R^\mu 
\Big\}~. 
\label{Action2ndOrder} 
} 

\subsection{Two--point quantum vertices}

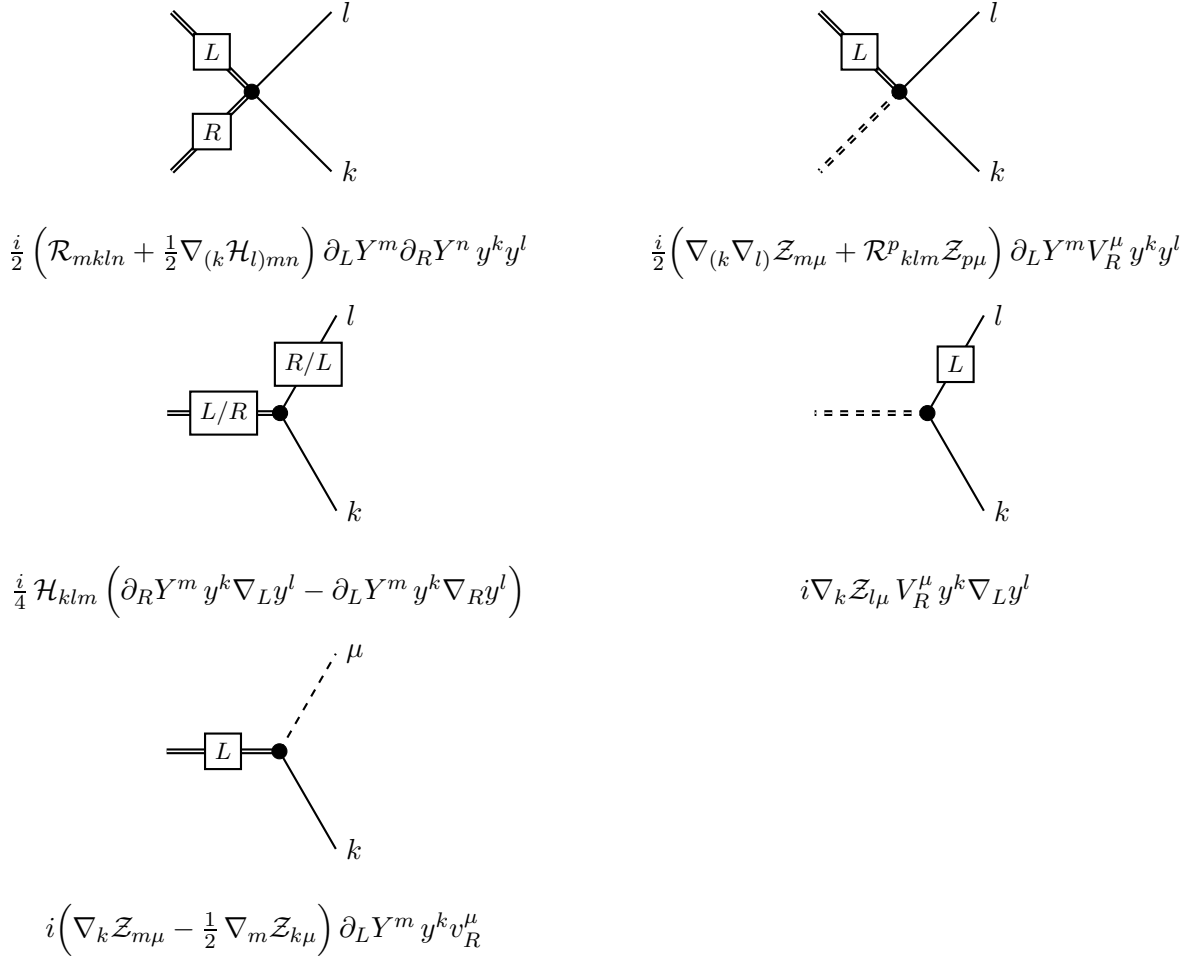
\begin{figure}[t]
\centering
\tabu{ccc}{ 
\begin{tikzpicture}
\draw[black, thick] (-45:0) -- (-45:1.5cm);
\draw[black, thick] (45:0) -- (45:1.5cm);
\draw[black, double, thick] (135:0) -- (135:1.5cm);
\draw[black, double, thick] (225:0) -- (225:1.5cm);
\filldraw[black] (0,0) circle (1mm);
\node[anchor=west] at (45:1.5) {$l$};
\node[anchor=west] at (-45:1.5) {$k$};
\node[draw, thick, fill=white, font=\footnotesize] at (135:0.75) {$L$};
\node[draw, thick, fill=white, font=\footnotesize] at (225:0.75) {$R$};
\end{tikzpicture}
& \qquad\qquad & 
\begin{tikzpicture}
\draw[black, thick] (-45:0) -- (-45:1.5cm);
\draw[black, thick] (45:0) -- (45:1.5cm);
\draw[black, double, thick] (135:0) -- (135:1.5cm);
\draw[black, double, thick, dashed] (225:0) -- (225:1.5cm);
\filldraw[black] (0,0) circle (1mm);
\node[anchor=west] at (45:1.5) {$l$};
\node[anchor=west] at (-45:1.5) {$k$};
\node[draw, thick, fill=white, font=\footnotesize] at (135:0.75) {$L$};
\end{tikzpicture}
\\[2ex] 
$\frac i2\, \Big( \cR_{mkln} + \frac{1}{2} \nabla_{(k} \cH_{l) mn} \Big) 
\,\dl Y^m \dr Y^n\, y^k y^l$
&&
$\frac{i}{2} \Big( \nabla_{(k} \nabla_{l)} \cZ_{m \mu} 
+ \cR^p{}_{klm} \cZ_{p \mu}  
\Big) \, \dl Y^m V_R^\mu\,  y^k y^l$
\\[2ex] 
%
\begin{tikzpicture}
\draw[black, thick, double] (180:0) -- (180:1.5cm);
\draw[black, thick] (-60:0) -- (-60:1.5cm);
\draw[black, thick] (60:0) -- (60:1.5cm);
\filldraw[black] (0,0) circle (1mm);
\node[anchor=west] at (60:1.5) {$l$};
\node[anchor=west] at (-60:1.5) {$k$};
\node[draw, thick, fill=white, font=\footnotesize] at (180:0.75) {$L/R$};
\node[draw, thick, fill=white, font=\footnotesize] at (60:0.75) {$R/L$};
\end{tikzpicture}
 &&
\begin{tikzpicture}
\draw[black, thick, double, dashed] (180:0) -- (180:1.5cm);
\draw[black, thick] (-60:0) -- (-60:1.5cm);
\draw[black, thick] (60:0) -- (60:1.5cm);
\filldraw[black] (0,0) circle (1mm);
\node[anchor=west] at (60:1.5) {$l$};
\node[anchor=west] at (-60:1.5) {$k$};
\node[draw, thick, fill=white, font=\footnotesize] at (60:0.75) {$L$};
\end{tikzpicture}
\\[2ex] 
$  \frac{i}{4}\, \cH_{klm}\, 
\Big( \partial_R Y^m \, y^k \nabla_L  y^l - \partial_L Y^m \, y^k \nabla_R  y^l \Big) $
&&
$i \nabla_k \cZ_{l\mu}\, V_R^\mu\, y^k \nabla_L y^l$
\\[2ex] 
\begin{tikzpicture}
\draw[black, double, thick] (-1.5,0) -- (0,0);
\draw[black, thick, dashed] (0:0) -- (60:1.5);
\draw[black, thick] (0:0) -- (-60:1.5);
\filldraw[black] (0,0) circle (1mm);
\node[anchor=west] at (60:1.5) {$\mu$};
\node[anchor=west] at (-60:1.5) {$k$};
\node[draw, thick, fill=white, font=\footnotesize] at (180:0.75) {$L$};
\end{tikzpicture}
\\[2ex] 
$i \Big( 
\nabla_k \cZ_{m \mu} - \frac 12\,  \nabla_m \cZ_{k \mu} 
\Big)\, \der_L Y^m \,   
y^k  v_R^\mu $
}
\caption{This table gives an overview of the vertices involving two quantum fields corresponding to the second and third lines of equation \eqref{Action2ndOrder}.
}
\label{fig:overvVertices}
\end{figure}


As should be clear from our background/quantum splitting described above, to represent the quantum corrections to the effective action we need to distinguish between quantum fields that can run around in loops and the background field insertions at the vertices. Therefore, we employ the following conventions to draw the Feynman diagrams: 

A single solid or dashed line ending at a vertex represents that at this vertex a quantum coordinate field $y$, or quantum Lagrange multiplier field $v_R$ couples, respectively. 
When on a solid line there is a box with $L$ or $R$ this means that on this field a left-- or right--moving covariant derivative, $\nabla_L$ or $\nabla_R$, acts, respectively. 
A solid line with an arrow pointing towards or away from the vertex denotes the insertion of an $b_R$-- or $c$--ghost, respectively.  
A double dashed line ending at a vertex denotes the insertion of a background Lagrange multiplier $V_R$. Similarly, a double solid line with a boxed $L$ or $R$ terminating at a vertex indicates that there a derivative of the background coordinate field, $\der_L Y$ or $\der_R Y$, is inserted, respectively. 
Finally, it should be stressed that each vertex is not a mere constant but rather a specific function of the background fields. 

Employing these conventions, the relevant vertices to construct all possible divergent one--loop diagrams are given in Table \ref{fig:overvVertices} ordered in the same way as the terms of the second and third lines of the expansion of the action to second order in the quantum fields \eqref{Action2ndOrder}.

\subsection{Progagators}
\label{sc:Props} 


The first line of \eqref{Action2ndOrder} encodes the kinetic terms of the quantum coordinate fields $y$ which partially mix with the quantum Lagrange multiplier $v_R$. In $d$--worldsheet dimensions with an IR--regulator $m^2$ these terms are given by
\equ{ \label{dextKin} 
S_\text{kin} = \gm^{d-2}
\int \d^d \gs\, \Big\{
\frac 14\, \hat\get^{ab}\, \nabla_a y^T \, \cG\, \nabla_b y 
+ \frac 14\, m^2\, y^T\, \cG\, y 
+ \frac 12\, v_R^T\, \cZ^T\, \nabla_L y 
- \frac 12\, y^T\, \cZ\, \nabla_L v_R
\Big\}~,
} 
where now $a,b=0,1,\ldots d-1$ and $\hat\get = \text{diag}(1,-1,\ldots, -1)$. Here we have introduced an arbitrary regularization scale $\gm$ to keep the mass dimensions as in two dimensions. 
%
%
We define the Fourier transform of covariant derivatives in $d$--worldsheet dimensions as 
 \equ{ \label{FourierTrans} 
 \gf(\gs) = 
 \int \dpp\, e^{i\, p \gs} \gf(p)~, 
 \qquad 
 \nabla_a \gf(\gs) = 
  \int \dpp\, e^{i\, p \gs} (ip)_a \gf(p)~, 
 }
for any covariant field $\gf$ such as $y$ or $v_R$. (This definition is compatible with the covariant Leibniz rule: 
$\der_a [\gf_1^T(\gs) \cG \gf_2(\gs)] = 
\nabla_a \gf_1^T(\gs) \cG \gf_2(\gs) + \gf_1^T(\gs) \cG  \nabla_a\gf_2(\gs)$.) 
Using this we can identify the inverse of the propagator $\gD$ for $y$ and $v_R$: 
\equ{ \label{invProp} 
S_\text{kin} = 
\frac 12\, \int \dpp\,
\begin{pmatrix} 
y^T & v_R^T
\end{pmatrix}(-p)\, 
\gD^{-1} \, 
\begin{pmatrix}
y \\ v_R
\end{pmatrix}(p)~, 
\quad 
\gD^{-1} = 
\left(\begin{array}{cc} 
\frac 12\, \cG\, (p^2+m^2) & -i\, \cZ\, p_L \\[1ex] i\, \cZ^T\, p_L & 0
\end{array}\right)~. 
}
Here we made the assumption that we are only interested in loop--momenta $p_a$ much larger than any momentum scale corresponding to the external background fields contained e.g.\ in $\cG$ and $\cZ$.  
%
%
By computing the inverse of \eqref{invProp} under the assumption that the metric $\cG$ is invertible, we can determine the propagator:
\equ{ \label{propagator} 
\gD = 
\begin{pmatrix}
{\G}_\perp^{-1}\, \frac{2}{p^2 + m^2} 
& 
{\cZ}_\parallel \,\frac{1}{ip_L}  
\\[1ex] 
- {\cZ}_\parallel^T\,  \frac{1}{ip_L} 
&  
- \left( \cZ^T \G^{-1} \cZ \right)^{-1}\, \frac{p^2 + m^2}{2\, p_L^2}
\end{pmatrix}~,  
}
where we have made use of the notation defined in \eqref{Uinvariants}. The different components of this combined propagator for $y$ and $v_R$ are visualized as follows: 
\begin{subequations} 
\equa{ \label{Propyy} 
\langle y^m y^n \rangle 
\quad = \quad 
\begin{tikzpicture}
\draw[black, thick] (0,0) -- (2.5,0);
\node[anchor=east] at (0,0.3) {$m$};
\node[anchor=west] at (2.5,0.3) {$n$};
\end{tikzpicture} 
\quad &= \quad 
-i\, ({\G}_\perp^{-1})^{mn}\, \frac{2}{p^2 + m^2}~, 
\\[1ex] \label{PropyvR} 
\langle y^m v_R^\nu \rangle
\quad = \quad 
\begin{tikzpicture}
\draw[black, thick] (0,0) -- (1.25,0);
\draw[black, thick, dashed] (1.25,0) -- (2.5,0);
\node[anchor=east] at (0,0.3) {$m$};
\node[anchor=west] at (2.5,0.3) {$\nu$};
\end{tikzpicture}
\quad &= \quad 
 -({Z_\parallel})^{m\nu}\,  \frac{1}{p_L}~, 
\\[1ex] 
\langle v_R^\mu v_R^\nu \rangle 
\quad = \quad 
\begin{tikzpicture}
\draw[black, thick, dashed] (0,0) -- (2.5,0);
\node[anchor=east] at (0,0.3) {$\mu~$};
\node[anchor=west] at (2.5,0.3) {$\nu$};
\end{tikzpicture}
\quad &= \quad 
 i\,  \Big( (\cZ^T \cG^{-1} \cZ )^{-1}\Big)^{\mu\nu}\,
 \frac{p^2 + m^2}{2\, p_L^2} ~. 
} \label{ExplProps} 
\end{subequations} 

The appearance of the propagator $\langle v_R^\mu v_R^\nu \rangle$ is somewhat surprising since the kinetic operator \eqref{invProp} has a zero for its $\gm\gn$--components. Moreover, the form of this propagator is rather non--standard. Fortunately, it turns out that it never contributes to any of the divergent diagrams of interest in this paper. 

\subsection{Ghost sector} 
\label{sc:Ghosts}

%
 \begin{figure}[t]
\begin{center}
\tabu{ccc}{
\raisebox{7.7ex}{
\begin{tikzpicture}
\draw[black, dotted, very thick] (0,0) -- (2.5,0);
\end{tikzpicture}
}
&\qquad\qquad\qquad   & 
\begin{tikzpicture}
\draw[black, double, thick] (-1.5,0) -- (0,0);
\draw[black, dotted, very thick] (0:0) -- (60:1.5);
\draw[black, dotted, very thick] (0:0) -- (-60:1.5);
\filldraw[black] (0,0) circle (1mm);
\node[draw, thick, fill=white, font=\footnotesize] at (180:0.75) {$L$};
\end{tikzpicture}
}
\end{center}
  \caption{Ghost propagator and its fundamental vertex are displayed.}
  \label{fig:ghostVertex}
\end{figure}
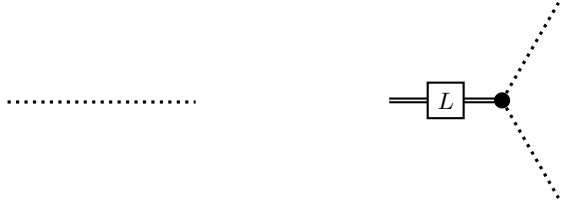
 

The Feynman rules discussed so far ignored the ghosts present in the action \eqref{BRSTact}. Contrary to the coordinate fields, the ghosts only appear quadratically in the action, hence we do not need to set up a covariant expansion for them; their path integral simply gives a formal fermionic determinant. To be able to evaluate this determinant perturbatively, we first cast the ghost action in a symmetric form, 
\equ{
S_\text{gh} = \int \dsigma\, \Big\{ 
\nabla_L c_\ga\, B_R^\ga  
+ c_\ga\, (A_{L}){}^\ga{}_\gb\, B_R^\gb
\Big\}~,
}
involving a $\go$--covariant derivative $\nabla_a$ which can be defined in a similar way as the $\rho$--derivative in \eqref{rhoCovDer}. Since the $c$-- and the $b_R$--ghosts carry different types of indices transforming under the $\go$-- and $\gr$--transformations, respectively, we have defined 
\equ{
B_R^\ga = N^\ga{}_\gm\, b_R^\gm~. 
}
Here we have used the matrix $N$ introduced in \eqref{fullyFixed} which is invertible provided that the constraint \eqref{GFfunct} fixes all the gauge symmetries. This then allows to introduce a ``connection'' $A_L$ for the ghost sector
\equ{
(A_L) ^\ga{}_\gb = 
\big( 
\cK^{p\ga}\, \nabla_p \cZ_{m\gm} 
+ \nabla_m \cK^{p\ga}\, \cZ_{p\gm}
\big)\, (N^{-1})^\gm{}_\gb~. 
}
Notice that the connection reflects the fact that the ghost sector is chiral, i.e.\ $A_L$ exists but $A_R$ does not. 
 

The extension of chirality to $d$--dimensions in dimensional regularization is a bit subtle. As far as the underlying gamma algebra is concerned, following \cite{tHooft.1972} we take the chirality operator to be defined as in two dimensions, i.e.\ $\tgg = \gg^0 \gg^1$. In addition, we extend the fermions by (unphysical) components of the opposite chirality as the ghost sector actually possesses; in two dimensions the physical components are identified as 
\equ{ 
\frac{1 - \tgg}2\, \gps = 
\frac 1{\sqrt 2}
\left(\begin{array}{c}0 \\b - i\, c\end{array}\right)~. 
}
However, like the chirality operator, the connection $A_a = (\hat e_L)_a A_L$ is taken to be a strictly two--dimensional object, since it depends on the background fields only, by introducing the unit vectors 
\equ{ \label{unitvec} 
\hat e_{L/R} = \frac 1{\sqrt 2}\big(1, \pm 1, 0,\ldots,0\big)~,
\qquad 
\hat e_{L/R}^T\,  \hat\get\, \hat e_{L/R} = 0~, 
\quad 
\hat e_{L/R}^T \, \hat\get\,  \hat e_{R/L} = 1~, 
} 
pointing in the left-- or right--directions, respectively. The ghost action extended to $d$ dimensions then takes the form: 
\equ{ \label{dextGhosts}
S_\text{gh} = \frac {\gm^{d-2}}{\sqrt 2} \int \d^d\gs\, 
\bgps\, \Big( 
\gg^a\, \nabla_a + m 
+ \gg^a\, A_a\, \frac {1-\tgg}2
\Big)\, \gps~. 
}
The corresponding propagator and vertex are depicted in \autoref{fig:ghostVertex}.

 \section{One loop renormalization}
 \label{sc:oneloop}

 \subsection{Renormalization of the kinetic term}
 \label{sc:renormKin} 
 
%
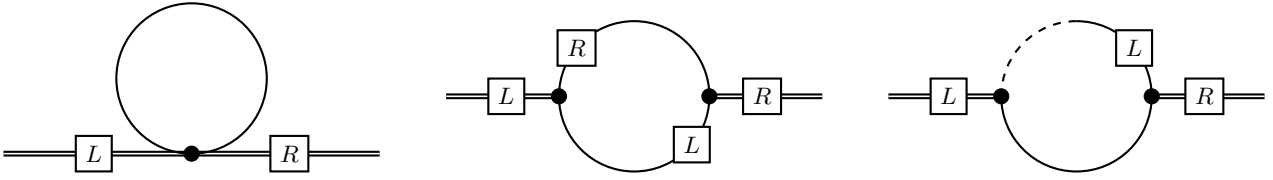
\begin{figure}[t]
\centering
\tabu{ccc}{ 
\begin{tikzpicture}
\draw[black, double, thick] (-2.5,-1) -- (2.5,-1);
\draw[black, thick] (0,0) circle (1);
\filldraw[black] (0,-1) circle (1mm);
\node[draw, thick, fill=white, font=\footnotesize] at (-1.3,-1) {$L$};
\node[draw, thick, fill=white, font=\footnotesize] at (1.3,-1) {$R$};
\end{tikzpicture}
~&~
\begin{tikzpicture}
\draw[black, double, thick] (-2.5,0) -- (-1,0);
\draw[black, thick] (0,0) circle (1);
\draw[black, double, thick] (1,0) -- (2.5,0);
\filldraw[black] (-1,0) circle (1mm);
\filldraw[black] (1,0) circle (1mm);
\node[draw, thick, fill=white, font=\footnotesize] at (-1.7,0) {$L$};
\node[draw, thick, fill=white, font=\footnotesize] at (1.7,0) {$R$};
\node[draw, thick, fill=white, font=\footnotesize] at (-40:1) {$L$};
\node[draw, thick, fill=white, font=\footnotesize] at (140:1) {$R$};
\end{tikzpicture}
~&~
\begin{tikzpicture}
\draw[black, double, thick] (-2.5,0) -- (-1,0);
\draw[black, thick,dashed] (0,1) arc (90:180:1);
\draw[black, thick] (1,0) arc (0:-180:1);
\draw[black, thick] (1,0) arc (0:90:1);
\draw[black, double, thick] (1,0) -- (2.5,0);
\filldraw[black] (-1,0) circle (1mm);
\filldraw[black] (1,0) circle (1mm);
\node[draw, thick, fill=white, font=\footnotesize] at (-1.7,0) {$L$};
\node[draw, thick, fill=white, font=\footnotesize] at (1.7,0) {$R$};
\node[draw, thick, fill=white, font=\footnotesize] at (40:1) {$L$};
\end{tikzpicture}
}
\caption{Diagrams for the renormalization of the kinetic term.} 
\label{fig:KinRenorm}
\end{figure}

To obtain the divergent contribution of the kinetic terms to the effective
action, all possible diagrams that can be composed of the propagators \eqref{ExplProps} and the vertices given in \autoref{fig:overvVertices} that include background contributions proportional to $\dl Y^m \dr Y^n$ have to be considered. Collectively, these diagrams, which are displayed schematically in \autoref{fig:KinRenorm}, lead to 
%
%
\equa{ 
\Gamma_\mathrm{kin} = 
I_1 \int \dsigma\, \Big\{ &  
\Big( 
\cR_{ijkl} + \frac{1}{2} \nabla_i \cH_{l j k}
\Big) \, (\cG_\perp^{-1})^{i l} 
- \frac{1}{4} \cH_{p m j} \cH_{n q k}\, (\cG_\perp^{-1})^{m n} (\cG_\perp^{-1})^{p q}
\non \\[1ex]  
& + \cH_{p m k}\, 
\Big( 
\nabla_q \cZ_{j \nu} - \frac{1}{2} \nabla_j \cZ_{q \nu} 
\Big)\,  (\cG_\perp^{-1})^{p q}\,  ({\cZ}_\parallel)^{m \nu} 
\Big\}\,  \dl Y^j \dr Y^k~, 
\label{KinRenorm}
}
where $I_1$ is the divergent integral~\eqref{logdiv} defined in Appendix~\ref{sc:DimReg}. Let us briefly explain how the various contributions arise: 

%
The first contribution corresponds to the first diagram shown in~\autoref{fig:KinRenorm}. Given that this diagram has the topology of a tadpole graph, it is proportional to the integral $I_1$. The (non--standard) normalization of the propagator \eqref{Propyy} is compensated by the factor of $1/2$ in front of the $\dl Y \dr Y y y$--vertex, see~\autoref{fig:overvVertices}. Similarly, the detailed tensor structure can be verified. 

%
The next divergent contribution to the effective action corresponds to the second diagram in~\autoref{fig:KinRenorm}. In fact this diagram corresponds to four contributions depending on which of the two internal lines the left-- and right--derivatives, indicated by a boxed $L/R$, act. These derivatives can be written $\nabla_{L/R} = \hat e^a_{L/R}\cdot ip_a$ where the unit vectors $\hat e_{L,R}$ were introduced in~\eqref{unitvec}. The divergent part of each of these contributions turns out to be proportional to the tensor integral $J_{ab}(k)$, given in~\eqref{momentumtensorintegral}, contracted with $\hat e_L^a \hat e_R^b$. The tensor structure can be directly read off from the corresponding vertex in~\autoref{fig:overvVertices} and the propagator \eqref{Propyy}. The normalization of this contribution arises as follows: As observed above, this diagram really corresponds to four contributions each of which involves two identical vertices each equipped with a factor of $1/4$ and two propagators with a non--standard factor $2$, therefore, we have: $4\cdot \frac 12 \cdot (\frac 14)^2 \cdot 2^2 = \frac 12$. As explained in Appendix~\ref{sc:DimReg} the divergent part of $J_{ab}(k)$ equals $\frac 12\, \hat\get_{ab}\, I_1$. Hence, the contraction with $\hat e_L^a \hat e_R^b$ gives $\hat e_L^T \hat\get \hat e_R =1$ (see \eqref{unitvec}), so that the overall factor equals $1/4$ for the second contribution. 

%
The last divergent contribution is due to the third diagram of 
\autoref{fig:KinRenorm}. Here the effect of $v_R$ becomes manifest as this diagram involves the propagator \eqref{PropyvR} that mixes the coordinate fields and $v_R$. In this diagram there are two options to place the left--derivative on the internal lines. In momentum space this derivative gives an internal momentum factor $p_L$ (up to finite contributions) which is cancelled by the $1/p_L$ factor in the mixed propagator~\eqref{PropyvR}. Consequently, these two options give opposite contributions, but given that $\cH$ is totally anti--symmetric they add up. Hence, the normalization factor of $1/4$ of the $\cH$--vertex is compensated by a factor of two due to the two possible placements of the $\nabla_L$ derivative and the non--standard normalization of the propagator \eqref{Propyy}.

 \subsection{Renormalization of the constraint} 
 \label{sc:renormConst} 
 
%
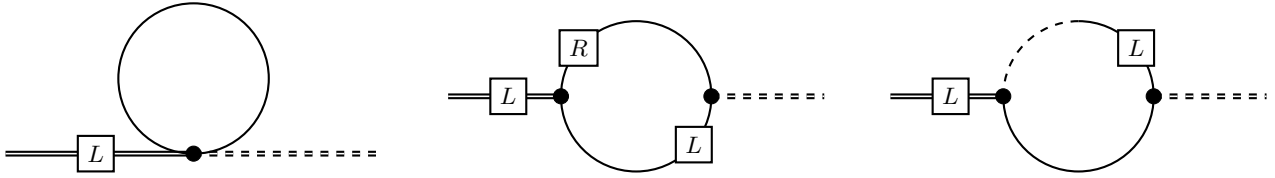
\begin{figure}[t]
\centering
\tabu{ccc}{ 
\begin{tikzpicture}
\draw[black, double, thick] (-2.5,-1) -- (0,-1);
\draw[black, double, thick,dashed] (0,-1) -- (2.5,-1);
\draw[black, thick] (0,0) circle (1);
\filldraw[black] (0,-1) circle (1mm);
\node[draw, thick, fill=white, font=\footnotesize] at (-1.3,-1) {$L$};
\end{tikzpicture}
~&~
\begin{tikzpicture}
\draw[black, double, thick] (-2.5,0) -- (-1,0);
\draw[black, thick] (0,0) circle (1);
\draw[black, double, thick, dashed] (1,0) -- (2.5,0);
\filldraw[black] (-1,0) circle (1mm);
\filldraw[black] (1,0) circle (1mm);
\node[draw, thick, fill=white, font=\footnotesize] at (-1.7,0) {$L$};
\node[draw, thick, fill=white, font=\footnotesize] at (-40:1) {$L$};
\node[draw, thick, fill=white, font=\footnotesize] at (140:1) {$R$};
\end{tikzpicture}
 ~&~ 
 \begin{tikzpicture}
\draw[black, double, thick] (-2.5,0) -- (-1,0);
\draw[black, thick,dashed] (0,1) arc (90:180:1);
\draw[black, thick] (1,0) arc (0:-180:1);
\draw[black, thick] (1,0) arc (0:90:1);
\draw[black, double, thick, dashed] (1,0) -- (2.5,0);
\filldraw[black] (-1,0) circle (1mm);
\filldraw[black] (1,0) circle (1mm);
\node[draw, thick, fill=white, font=\footnotesize] at (-1.7,0) {$L$};
\node[draw, thick, fill=white, font=\footnotesize] at (40:1) {$L$};
\end{tikzpicture}
 }
\caption{Diagrams for the renormalization of the constraint.}
\label{fig:ConstRenorm}
\end{figure}

To determine the divergent contribution to the effective action corresponding to the constraint one considers all diagrams proportional to $\dl Y V_R$; these are depicted in \autoref{fig:ConstRenorm}. The total divergent contribution to the constraint term in~\eqref{eq:act} at one loop level is given by
\equa{ 
\Gamma_\mathrm{con} 
= I_1\,  \int \dsigma\,\Big\{ & 
 \Big( 
\nabla_{(i} \nabla_{j)} \cZ_{k \nu} + \cR^{m}{}_{ijk} \cZ_{m \nu}  
\Big) \, (\cG_\perp^{-1})^{ij}
+ \cH_{p m k}\, \nabla_n \cZ_{q \nu}\, 
({\cG}_\perp^{-1})^{p q}  ({\cG}_\perp^{-1})^{m n}
\non \\[1ex]  &
- 4\, \nabla_{[p} \cZ_{m] \nu}\, 
 \Big( 
 \nabla_q \cZ_{k \mu} - \frac{1}{2} \nabla_k \cZ_{q \mu}
\Big) \, ({\cG}_\perp^{-1})^{p q} ({\cZ}_\parallel)^{m \mu}
  \Big\}\, \dl  Y^k \,V_R^\nu~. 
\label{ConstRenorm}
}
The constraint renormalization diagrams in \autoref{fig:ConstRenorm} are evaluated in a similar fashion as the diagrams in \autoref{fig:KinRenorm} for the kinetic renormalization using the vertices given in \autoref{fig:overvVertices} and the propagators \eqref{ExplProps}. Therefore we only focus here on the fundamental difference compared to the kinetic renormalization discussion in the previous Subsection. The only true difference appears in the third diagram of \autoref{fig:ConstRenorm}: As in the expression corresponding to the third diagram in \autoref{fig:KinRenorm} for the kinetic renormalization one finds two contributions with opposite sign. But in the present case, the vertex is not anti--symmetric itself, since it contains $\nabla_p \cZ_{m\nu}$. Therefore, the opposite sign contributions lead to an anti--symmetrization of the indices $p$ and $m$ as indicated in the third contribution in \eqref{ConstRenorm}.

\subsection{Absence of renormalization due to ghosts}
\label{sc:NoGhosts}

Even though the ghosts are very important for the internal consistency of the doubled worldsheet theory considered in this work, as we argue in this Subsection that they do not contribute to the renormalization of the constraint and kinetic terms at one loop. That they cannot renormalize the gauge fixing constraint term is obvious since there is simply no direct coupling of the ghosts to the Lagrange multiplier field $V_R$. 

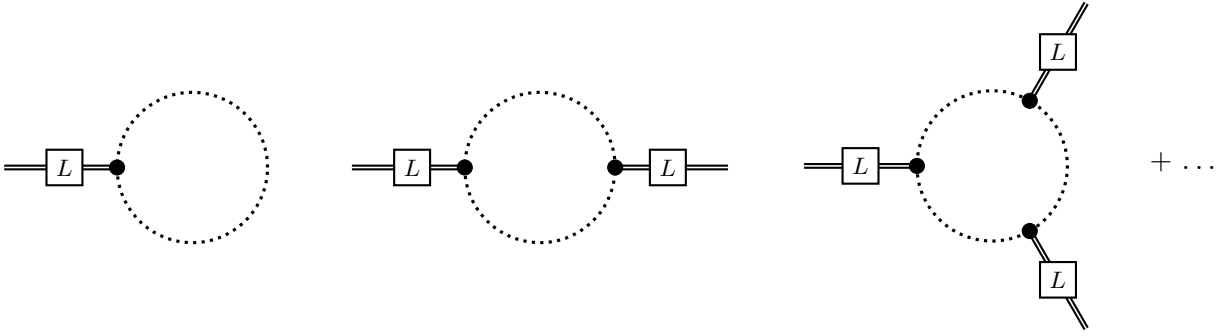
\begin{figure}[t]
\centering
\tabu{cccc}{
\raisebox{7ex}{
\begin{tikzpicture}
\draw[black, double, thick] (-1.5,0) -- (0,0);
\draw[black, dotted, very thick] (1,0) circle (1cm);
\filldraw[black] (0,0) circle (1mm);
\node[draw, thick, fill=white, font=\footnotesize] at (-0.7,0) {$L$};
\end{tikzpicture}
}
~&~
\raisebox{7ex}{
\begin{tikzpicture}
\draw[black, double, thick] (-2.5,0) -- (-1,0);
\draw[black, dotted, very thick] (0:1) arc (0:180:1);
\draw[black, dotted, very thick] (180:1) arc (180:360:1);
\draw[black, double, thick] (1,0) -- (2.5,0);
\filldraw[black] (-1,0) circle (1mm);
\filldraw[black] (1,0) circle (1mm);
\node[draw, thick, fill=white, font=\footnotesize] at (-1.7,0) {$L$};
\node[draw, thick, fill=white, font=\footnotesize] at (1.7,0) {$L$};
\end{tikzpicture}
}
~&~
\begin{tikzpicture}
\draw[black, double, thick] (0:-2.5) -- (0:-1);
\draw[black, double, thick] (120:-2.5) -- (120:-1);
\draw[black, double, thick] (240:-2.5) -- (240:-1);
\draw[black, dotted, very thick] (-60:1) arc (-60:60:1);
\draw[black, dotted, very thick] (60:1) arc (60:180:1);
\draw[black, dotted, very thick] (180:1) arc (180:300:1);
\filldraw[black] (0:-1) circle (1mm);
\filldraw[black] (120:-1) circle (1mm);
\filldraw[black] (240:-1) circle (1mm);
\node[draw, thick, fill=white, font=\footnotesize] at (180:1.75) {$L$};
\node[draw, thick, fill=white, font=\footnotesize] at (60:1.75) {$L$};
\node[draw, thick, fill=white, font=\footnotesize] at (-60:1.75) {$L$};
\end{tikzpicture}
~&~ 
\raisebox{13ex}{+ \ldots} 
}
\caption{The diagrammatic expansion of the fermionic ghost determinant (to the third order).}
\label{fig:ghostDiagrams}
\end{figure}

To understand that there is also no renormalization induced by the ghost fields to the kinetic terms of the doubled coordinate fields $Y$ is a bit more involved: As there is just one vertex involving ghost fields, see \autoref{fig:ghostVertex}, all one--loop diagrams that arise from expanding the formal fermionic ghost determinant have the same structure. The diagrams corresponding to this expansion are depicted in \autoref{fig:ghostDiagrams}. Since the ghost vertex involves the connection $A_L$ only and our regularization procedure preserves two dimensional Lorentz invariance at least, the diagram with $n$ background insertions will be proportional to $(\dr A_L)^n$. 

%
Furthermore, for the renormalization only the first two diagrams are relevant, since all other diagrams are finite. For the second diagram in \autoref{fig:ghostDiagrams} this would mean that the theory should have a divergence proportional to $(\dr A_L)^2$. However, by conformal invariance such a term cannot be present in the bare action, hence the divergence has to be absent. This can be verified explicitly: The second diagram is proportional to the integral $J_{ab}$ defined in \eqref{momentumtensorintegral} of Appendix~\ref{sc:DimReg}, its divergent part being proportional to $\hat\get_{ab}$. Contracting this with the $A_a$ gives zero by~\eqref{unitvec}. Hence either by formal arguments or by an explicit computation we conclude that the second diagram of \autoref{fig:ghostDiagrams} does not give a divergent contribution. 

%
Therefore, only the first graph in \autoref{fig:ghostDiagrams} can potentially lead to a renormalization of the theory. By the same argument as above, we conclude that this tadpole graph is proportional to $\dr A_L$. But this then just gives a total derivative in the effective quantum action and hence is irrelevant. 
%

 \subsection{Weyl invariance at the one--loop level}
 \label{sc:TSeom} 

%
Having determined all divergent contributions to the effective action we are now in the position to determine the target space equations of motion by requiring that the renormalized theory is invariant under Weyl transformations. To this end, we now consider the theory in conformal gauge instead of Minkowski gauge as before, i.e.\ we take the worldsheet metric $\gg'(\gs)$ to be
\equ{
\gg'(\gs) = e^{2 \gvf(\gs)}\,  \gg 
}
with a conformal factor $\gvf(\gs)$ and $\gg = \text{diag}(1,-1)$ denoting the flat Minkowski metric. As this worldsheet metric is related to the one used before by a Weyl transformation, this allows us to derive conditions for conformal invariance. 

%
For a non--Minkowski metric $\gg'$ one needs to include the Einstein--Hilbert action,  
\equ{ 
S_\text{EH} = \int \d^2\gs\, \sqrt{\gg'} R(\gg')\, \gF(Y)~, 
}
on the worldsheet involving the dilaton $\gF$. 
Using how the Ricci scalar transforms under a conformal transformation $\gg \to \gg' = e^{2 \gvf} \gg$,
\equ{
\sqrt{\gg'} R({\gg'}) = \sqrt{\gg} \left[ R(\gg) - 2 (d-1) \, \gg^{a b} \, \nabla_a \nabla_b \gvf - (d-2)(d-1) \, \gg^{ab} \, \nabla_a \gvf \nabla_b \gvf \right]~, 
}
we find
\equ{
\sqrt{\gg'} R({\gg')} = - 4 \dl \dr \gvf~, 
}
since $\gg$ is in Minkowski gauge. Thus the full action in conformal gauge is given by
\equ{ \label{eq:actConf}
S_\gvf  = \frac{1}{\ga'} \int \dsigma \Big( \frac 12\, \dl Y^T \,\cE\,\dr Y + \cW_L\, V_R - 4\ga'  \, \gvf \, \dl \dr \gF \, \Big)
}
after integrating by parts twice in the dilaton $\gF(Y)$ term.

%
Now considering \eqref{eq:actConf} in $d= 2 - 2 \ge$ dimensions to employ dimensional regularization, because of
\equ{
\sqrt{\gg'} = e^{(2-2 \ge) \gvf} \sqrt{\gg} \quad \text{and} \quad {\gg'}^{ab} = e^{- 2 \gvf} {\gg}^{ab}~, 
}
one finds that the factors multiplying $\sqrt{\gg'}$ and ${\gg'}^{ab}$ do not cancel exactly but a factor $e^{- 2 \ge \gvf}$ remains in front of the Lagrangian density. Thus in conformal gauge all vertices carry a factor of $e^{- 2 \ge \gvf}$ and all propagators are multiplied by $e^{2 \ge \gvf}$. However, as for all planar diagrams with $V$ vertices, $F$ faces (including the outer one) and $P$ propagators one has $V + F - P = 2$. It follows that in particular every one loop graph satisifies $V = P$. Thus these factors cancel for all diagrams that contribute to the divergent contribution.

%
So for infinitesimal $\gvf$ the renormalized effective action in conformal gauge is given by
\equa{
\Gamma_\gvf &= \frac{1}{\ga'} \int \dsigma \left[ e^{- 2 \ge \gvf} \left( \cL - \ga' 4 \, \gvf \, \dl \dr \gF - \frac{1}{4 \gp \ge} \cL_{\text{ct}} \right) + \ga' \cL_{\text{div}} \right] \non \\
&= \frac{1}{2 \gp \ga'} \int \dsigma \, \left[ 2 \gp \cL - \gvf \, \left( \cL_{\text{ct}} + 2 \ga' \dl \dr \gF \right) + \frac{1}{2 \ge} \left(\cL_{\text{ct}} + \ga' \cL_{\text{div}} \right) \right]
}
where $\cL$, $\cL_{\text{ct}} = \ga' \cL_{\text{div}}$ and $\cL_{\text{div}}$ denotes the Lagrangian, counterterm Lagrangian and divergent contribution in Mikowski gauge. Thus, finiteness of the theory requires $\cL_{\text{ct}} + \ga' \cL_{\text{div}} = 0$. This may now be used to turn the condition $\cL_{\text{ct}} - 2 \ga' \dl \dr \gF = 0$ that ensures conformal invariance into the form
\equ{
\cL_{\text{div}} - 2 \dl \dr \gF = 0~.
}
%
%
By the classical field equation \eqref{eq:classEOM} one obtains
\equa{  \label{eq:covn}
\dl \dr \gF &= \nabla_m \nabla_n \gF\, \dl Y^m \dr Y^n + \nabla_m\, \gF \nabla_L \nabla_R Y^m  
\\
&= \nabla_n \nabla_m \gF\, \dl Y^m \dr Y^n - \frac{1}{2} \nabla_m \gF\, \cG^{m n}\, \cH_{n a b}\, \dl Y^a \dr Y^b 
\non \\ 
&\quad + \partial_m \gF \cG^{m n}\, \Big(\cZ_{n \nu}\, \dl \vr^\nu + 2 \cD_{[n} \cZ_{p] \nu}\, \dl Y^p \vr^\nu \Big)~. 
\non 
}

This gives us three conditions for conformal invariance corresponding to three target space equations of motion. The first terms have the background field structure of the divergent contribution corresponding to the kinetic term, i.e.\  $\dl Y^j \dr Y^k$, and can thus be combined with this divergent contribution to give
\equa{
\Big( 
\cR_{ijkl} &+ \frac{1}{2} \nabla_i \cH_{l j k}
\Big) \, (\cG_\perp^{-1})^{i l} 
- \frac{1}{4} \cH_{p m j} \cH_{n q k}\, (\cG_\perp^{-1})^{m n} (\cG_\perp^{-1})^{p q}
\non \\[1ex]  
&+ \cH_{p m k}\, 
\Big( 
\nabla_q \cZ_{j \nu} - \frac{1}{2} \nabla_j \cZ_{q \nu} 
\Big)\,  (\cG_\perp^{-1})^{p q}\,  ({\cZ}_\parallel)^{m \nu} 
-2 \, \nabla_j \nabla_k \gF + \nabla_m \gF \, \cG^{m n} \, \cH_{n j k}
 = 0~.
\label{eq:eomKin}
}
The background field structure of the third term matches the one of the constraint's divergent contribution proportional to $\dl  Y^k \,V_R^\nu$. Thus they can be combined into the condition
\equa{
 \Big( 
\nabla_{(i} \nabla_{j)} \cZ_{k \nu} &+ \cR^{m}{}_{ijk} \cZ_{m \nu}  
\Big) \, (\cG_\perp^{-1})^{ij}
+ \cH_{p m k}\, \nabla_n \cZ_{q \nu}\, 
({\cG}_\perp^{-1})^{p q}  ({\cG}_\perp^{-1})^{m n}
\non \\[1ex]  
&- 4\, \nabla_{[p} \cZ_{m] \nu}\, 
 \Big( 
 \nabla_q \cZ_{k \mu} - \frac{1}{2} \nabla_k \cZ_{q \mu}
\Big) \, ({\cG}_\perp^{-1})^{p q} ({\cZ}_\parallel)^{m \mu} 
 -4 \, \partial_m \gF \, \cG^{m n} \, \partial_{[n} \cZ_{k] \nu}
   = 0~.
\label{eq:eomCon}
}
The final term is covariant even though an ordinary derivative appears here because of anti--symmetry. Lastly, the fourth term's background field structure $\dl \vr^\nu$ does not appear in $\cL_{\text{ct}}$. Therefore, it has to vanish by itself: 
\equ{ \label{eq:eomDil}
\partial_m \gF \, (\cG^{-1})^{m n} \, \cZ_{n \nu} = 0~.
}
As a cross check we confirmed that the equation~\eqref{eq:eomKin} reduces to the standard equations of motions of the metric and $B$-field when we use the standard sigma model form.

 \section{Target space interpretation}
 \label{sc:TSdiscussion} 
    
In the entire paper we have primarily considered the worldsheet perspective of the doubled sigma model. In this section we interpret various aspects of our doubled worldsheet theory from the target space point of view. A summary of both worldsheet and target space interpretations of the building blocks of the doubled worldsheet theory has been collected in~\autoref{tb:DoubledObjects} of the introduction. 

\begin{figure}
\begin{equation*}
\xymatrix{
*++[F-,]{Y^m(\sigma)} \ar[rrrr]^-{\text{Gauging~\eqref{gaugesym}}} & & & & *++[F-,]{X^\mu(\sigma)} \\
*++[F-,]{\mathcal{E}_{mn}(Y)} \ar[rr]^-{\text{Projection~\eqref{ProjectionEqs}}} & \hspace{2cm} & *++[F-,]{\tilde{\mathcal{E}}_{mn}(Y)} \ar[rr]^-{\text{Killing equation~\eqref{KillingEqs}}} & \hspace{2cm} & *++[F-,]{\tilde{\mathcal{E}}_{mn}(X) \cong E_{\mu\nu}(X)}
}
\end{equation*}
\caption{\label{fg:Reduction}
This figures indicates the reduction of the number of degrees of freedom of the doubled coordinate fields $Y^m$ on the worldsheet and the double metric and anti--symmetric tensor field contained in $\cE_{mn}$ of the doubled geometry.}
\end{figure}
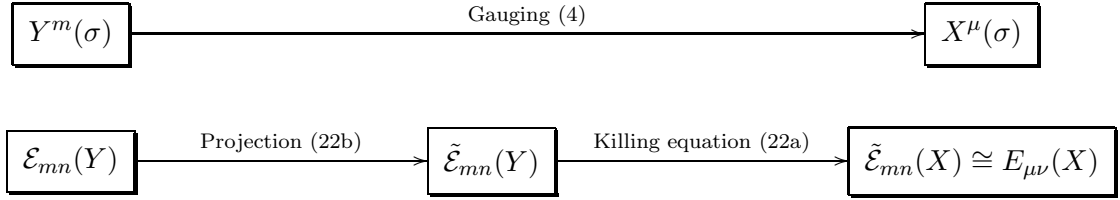

%
The starting point of our doubled sigma model were the $2D$ coordinate fields $Y$ subject to the gauge transformations~\eqref{gaugesym}. In target space descriptions with doubled coordinates one formulates the whole theory as if it had $2D$ coordinates, though at some point one enforces that only $D$ of them are physical. For example, in double field theory~\cite{hz09} one has to enforce the so--called strong constraint at various stages. Hence, as schematically indicated in~\autoref{fg:Reduction}, one can view the gauge transformations~\eqref{gaugesym} as the worldsheet manifestation of this reduction of the number of target space coordinates. 

%
In double field theory the solution to the strong constraint is far from being unique: One can define various so--called polarizations to solve it. In our worldsheet description things are quite similar: The gauge transformations~\eqref{gaugesym} are part of the very definition of the theory, while the choice of the gauge fixing condition~\eqref{GFfunct} and, in particular, of the the matrix $\cZ$, is essentially be chosen at will as long as $\cZ^T \cK$ is invertible. As we have seen, the appearance of many objects within our theory depends on the choice of $\cZ$. The form of the propagators for the quantum perturbations of the coordinate fields shows that this extends to the quantum theory as well: They involve the matrix $\cG^{-1}_\perp$ which is the inverse of the doubled metric $\cG$ projected in the directions perpendicular to $\cZ$, hence, the quantum theory ``knows'' that only $D$ of the $2D$ coordinate fields propagate. 

%
As usual, the functions in the worldsheet action are interpreted as target space fields. E.g.\ $\cG$ and $\cC$ are interpreted as the metric and anti--symmetric tensor field on the doubled target space, respectively. Given that $\cZ$ parameterizes the gauge fixing and hence is not part of the physical definition of the theory, it should not be interpreted as a dynamical target space field. 

%
On the other hand, in the doubling process from $E = g+b$ to $\cE = \cG+\cC$ many -- in principle -- arbitrary choices are made. As we showed in Table~\ref{tb:StForms}, we can bring $\cG$ e.g.\ in the form of the $O(D,D)$ invariant metric $\get$ or the $O(D,D)$ covariant generalized metric $\cH$. The $\cU$--transformation defined in~\eqref{Utrans} that relates these different descriptions is a local (i.e.\ $Y$ dependent) transformation in target space. 

%
Our worldsheet gauge theory provides a deeper insight in the origin of this target space gauge symmetry: When we want to describe the redundancy of the doubling of the coordinate fields at the level of the path integral we need to extend the gauge symmetry~\eqref{gaugesym} to full--fledged BRST transformations involving ghost fields. As we saw in Subsection~\ref{sc:BRST}, the precise form of the BRST transformation is not uniquely determined. The ambiguities that are visible even for the bosonic worldsheet fields are measured by the matrix $\cQ$ in the BRST transformations~\eqref{BRST_bR} and~\eqref{BRST_VR}. As usual, ambiguities in the worldsheet description lead to target space gauge symmetries. (Recall e.g.\ the fact that there is no preferred choice for the field basis leads to target space diffeomorphisms.) In the present case the ambiguities in the BRST symmetry on the worldsheet induce $\cU$--gauge transformations in target space. 

%
The analysis of the BRST transformations in Subsection~\ref{sc:BRST} showed that the worldsheet theory only possesses the gauge symmetry~\eqref{gaugesym} in the path integral provided that the functions $\cK$, that parameterize these gauge transformations, fulfill the projective Killing equations~\eqref{BRSTconditions}. As discussed in Subsection~\ref{sc:Stdouble}, the conditions~\eqref{ProjectionEqs} imply that we can bring $\cE$ to the standard sigma model form (in the $\cU$--transformation gauge $\cQ=0$). In other words, these projection conditions reduce the number of physical independent components of $\tilde\cE$ from $4D^2$ to the number $D^2$ of the standard sigma model. Moreover, the remaining terms in the Killing equation~\eqref{KillingEqs} then imply that these $D^2$ components are not a function of the dual coordinates. In this sense, the Killing equations and the strong constraint of double field theory seem to be closely related. In~\autoref{fg:Reduction} both this reduction of the number of components of $\cE$ and the restriction of its coordinate dependence are displayed.

%
In light of this fact, the target space meaning of equation~\eqref{ExpcE} becomes clear: It simply tells us that apart from the $D^2$ physical components in $\tilde\cE$ the other $3\, D^2$ components of $\cE$ are redundant. Indeed, the matrix functions $\cQ$ and $\cZ$ together have $3\,D^2$ independent components since both of them are $2D\times D$ matrices, but $D^2$ components can be removed from $\cZ$ using the $\rho$--transformation~\eqref{rhotrans}. In other words, the matrices $\cQ$ and $\cZ$ simply parameterize the non--physical components of the doubled metric $\cG$ and anti--symmetric tensor $\cC$. 

%
The above discussion shows that our worldsheet theory is  in the spirit of double field theory in quite a few respects. However, there are also some fundamental differences. For example, double field theory is invariant under so--called double field theory gauge transformations~\cite{hhz10b} 
\equ{ \label{DFTgaugeTrans}
\gd_\text{DFT} \cH_{mn} = 
\gx^p\, \cH_{mn,p} 
+ \cH_{mp}\, (\gx^p{}_{,n} - \gx_{n,}{}^{p}) 
+ (\gx^p{}_{,m} - \gx_{m,}{}^p) \, \cH_{pn}~, 
}
where the indices are raised/lowered using the $O(D,D)$ invariant metric $\get$. This transformation can be understood as the $2D$--Lie derivative, acting on the generalized metric $\cH$, made compatible with the condition that the generalized metric is itself an $O(D,D)$ element. If it acted as just the ordinary $2D$--dimensional generalization of the Lie--derivative, this transformation would be induced within the worldsheet theory as infinitesimal doubled diffeormorphisms~\eqref{doubleddiffs}, because the worldsheet derivatives of the coordinate fields, $\der_a Y$, naturally transform as covariant vectors in $2D$ dimensions. However, contrary to what is sometimes claimed in the literature~\cite{c11}, the additional contributions in the brackets in~\eqref{DFTgaugeTrans} can not be reproduced from a doubled worldsheet theory. Since this argument only relies on how worldsheet derivatives of coordinate fields transform, it applies to any doubled sigma model formulation on the worldsheet, including the non--Lorentz invariant worldsheet description of Tseytlin~\cite{t90,t91}. 

%
So far we only discussed the interpretation of the worldsheet theory in target space at the classical level. In Section~\ref{sc:oneloop} we have derived the conditions ensuring that the worldsheet theory is Weyl invariant at the one--loop level. For the conventional sigma model, the conditions of Weyl invariance on the worldsheet translate into the target space equations of motion in leading order in the $\ga'$--expansion. As the results, \eqref{eq:eomKin}--\eqref{eq:eomDil},  in Subsection~\ref{sc:TSeom} show, the interpretation of the Weyl invariance conditions for the doubled theory is a bit more subtle because one finds three rather than one equation. (We did not compute the renormalization of the Einstein--Hilbert term on the worldsheet itself, so we do not have access to the Dilaton equation of motion.) The first equation~\eqref{eq:eomKin} can be understood as a direct generalization of the standard beta function of the conventional sigma model. However, there are two crucial differences: i) There are additional terms due to the gauge fixing constraint. ii) Where in the standard case the inverse metric is contracted, giving e.g.\ the Ricci--tensor, now the projected inverse of the doubled metric, $\cG^{-1}_\perp$, appears. As we observed above, this is due to the fact that only $D$ quantum coordinate fields effectively propagate on the worldsheet and therefore contribute in loops. 

%
As we explained above, the projective Killing equations~\eqref{BRSTconditions} encode that the target space fields are effectively only functions of the original coordinates $X$. The constraint equation~\eqref{eq:eomDil} for the dilaton on the doubled target space leads to a similar conclusion. It tells us that $D$ combinations of doubled derivatives vanish on the dilaton. However, contrary to the Killing equation~\eqref{BRSTconditions}, in this equation not the Killing vectors $\cK$ but rather the gauge fixing function $\cZ$ appears. 

%
The target space equations of motion we derived are not identical those of double field theory even when we bring the worldsheet theory to the form in which the kinetic term involves the generalized metric. There seems to be a good reason for this discrepancy: As we explained above, the doubled worldsheet theory is incompatible with the double field theory gauge transformations. Given that these are a symmetry of the double field theory equations of motion, it is not to be expected that they are identical to the effective target space equation of motion derived from the doubled worldsheet theory. The reason, that our worldsheet formalism automatically has $2D$ diffeomorphisms build in, is that it is a complete off-shell realization; i.e.\ we do not distinguish between the zero modes and the full quantum fields.

\section{Outlook} 
\label{sc:outlook} 

%
To conclude, we discuss some questions which our work has left open and thereby give some suggestions for possible future research work. 

%
For the determination of the covariant Feynman rules we have tactically assumed that the doubled metric $\cG$ is invertible. In Section~\ref{sc:covFey} we argued that we can always use a specific $\cU$--transformation~\eqref{Utrans} to turn a non--invertible doubled metric, like the one naturally provided by the standard sigma model, into an invertible one. This leads to at least two questions: i) What happens if one decides not to use the $\cU$--transformations? ii) Since the precise form of the $\cU$--transformation is quite arbitrary, how does one see that the physical results do not depend on the $\cU$--gauge choice? 

As to the first question: Even though we prefer to refrain from giving details here, we have directly determined the propagators for the non--invertible doubled metric associated with the standard sigma model description. While $\cG^{-1}$ does not exist, one can still determine the propagators: There is an independent one for the quantum version of the coordinate fields $X$ and there are mixed propagators for the dual coordinates $\tX$ and $V_R$. Moreover, one could parameterize a $\cU$--transformation such that the original non--invertible metric is recovered in an specific limit. For many results one can confirm that this limit can be taken without any problems.

To the second question: In fact, the dependence on the $\cU$--gauge is less severe than one would naively expect. For example, we have seen that the propagators~\eqref{Propyy} are proportional to $\cG^{-1}_\perp$ defined in~\eqref{Uinvariants} which turns out to be inert under $\cU$--transformations. This is particularly helpful when applying the limiting procedure for non--invertible $\cG$ as described above. On the other hand, the conditions that guarantee Weyl invariance~\eqref{eq:eomKin}--\eqref{eq:eomDil} are not represented in terms of $\cU$--inert objects only. We have tried to construct combinations of these equations that are invariant under $\cU$--transformations, but unfortunately did not recover such rewritings.

%
One of the main results of this paper are the equations of motion in the doubled theory. An interesting but potentially complicated question is whether they can be derived from an action. Presumably, this will be an action formulated on the doubled target space. Although one could expect similarity to the formulation of double field theory, there are some indications that this theory and ours will not be identical: By working in the form where the worldsheet kinetic term is given by the generalized metric, we have tried to recover the double field theory equations of motion. However, even though we find many similar terms, it seems impossible to have a complete matching. This could be simply due to the fact that one compares the theories in $2D$ rather than $D$ dimensions, i.e.\ before additional constraints, like the strong constraint of double field theory or the dilaton equation \eqref{eq:eomDil} and the Killing equation \eqref{KillingEqs}, have been imposed. In other words, on the level of the doubled theories, our results and double field theory do not seem to be identical. Furthermore, as mentioned in Section \ref{sc:TSdiscussion}, the doubled worldsheet theory considered here is naturally invariant under $2D$--diffeomorphisms whereas double field theory is invariant under double field theory gauge transformations. Even though these are related, they are not the same.

%
Let us close with some rather general considerations about doubled worldsheet theories. The doubling discussed here is similar to the approaches of Tseytlin \cite{t90,t91} and Hull \cite{h06,Hull.2005} in that it is off-shell on the worldsheet. However, one may wonder whether an off--shell doubling is necessary at all. In a sense, the standard sigma model offers an on-shell variant: The left-- and right--moving coordinates are treated independently, both possessing zero modes from which target space coordinates $X$ and dual coordinates can be constructed. For example, the elaborate asymmetric orbifold constructions \cite{Condeescu:2012sp,Condeescu:2013yma} use the conventional worldsheet theory in the fermionic formulation without any off-shell doubling. Eventually, one could suspect that double field theory is more closely related to the on--shell left-- and right--moving zero modes of the standard sigma model coordinate fields than to the fully off--shell doubled coordinate fields $Y$. This might offer an alternative approach to realize double field theory within a worldsheet formalism.

\appendix 
\def\theequation{\thesection.\arabic{equation}} 
\setcounter{equation}{0}
 \section{Details of the covariant expansion}
 \label{sc:covExp} 
 
In this Appendix we briefly explain some of the computational steps involved in the method outlined in Subsection~\eqref{sc:CovSplitting} to obtain the expansion of the action~\eqref{eq:act} to second order in the covariant quantum fields $y$ and $v_R$ given in~\eqref{Action2ndOrder}. 

%
When working out the $s$--differentiation on the various terms in the action, one can replace all partial derivatives with the corresponding fully covariant ones introduced in Subsection \ref{sc:CovDers}. This is possible because the action itself is a scalar, hence all connection terms in the various covariant derivatives will cancel among each other. Concretely, for example, for the $s$--derivatives of the kinetic terms one obtains
$K(s) = \G_{m n}(s)\, \dl Y^m(s) \dr Y^n(s)$ we find: 
\equ{ 
\frac{\der}{\der s}\Big[ 
\G_{m n}(s)\, \dl Y^m(s) \dr Y^n(s)
\Big] = 
\G_{m n}(s) \Big( 
\nabla_L \dot Y^m(s) \dr Y^n(s) +  \dl Y^m(s) \nabla_R \dot Y^n(s) \Big)~;
}
where the additional term involving a covariant derivative on the metric vanishes using metric compatibility \eqref{MetricComp}. 
%
%
Applying a second $s$--derivative gives 
\equa{
\frac{\der^2}{\der s^2} \Big[ 
\G_{m n}(s)\, &  \dl Y^m(s) \dr Y^n(s)
\Big] =   
2\, \G_{m n}(s)\, \nabla_L \dot Y^m(s) \nabla_R \dot Y^n(s)  
 \non \\[1ex]  
& \qquad\qquad  
+ \G_{m n}(s)\, [\nabla_s, \nabla_L] \dot Y^m(s) \dr Y^n(s) 
+ \G_{m n}(s)\, \dl Y^m(s) [\nabla_s, \nabla_R] \dot Y^n(s) 
\non \\[2ex]  
= &\ 2\, \G_{m n}(s)\, \nabla_L \dot Y^m(s) \nabla_R \dot Y^n(s) 
+ 2 \, \cR_{ijkl}(s)\,  \dot Y^i(s) \dot Y^l(s) \, \dl Y^j(s) \dr Y^k(s)~.
}
The commutators can be introduced because the geodesic equation \eqref{eq:geodEq} gives: $\nabla_s \dot Y(s) = \nabla_s^2 Y(s) = 0$. The commutators of the covariant derivatives can then be replaced by curvature tensors \eqref{Curvature}. 


For the terms in the action involving the anti--symmetric tensor field $\cC$ in the doubled space, we first notice that after integrating by parts we have 
\equ{
\nabla_L \dot  Y(s)\, \mathcal{C}(s)\, \dr Y(s) 
+  \dl Y(s)\,  \mathcal{C}(s)\, \nabla_R \dot  Y(s) =  
-  \dot  Y(s)\, \nabla_L \mathcal{C}(s)\,  \dr Y(s)  
-  \dl Y(s) \, \nabla_R \mathcal{C}(s)\,  \dot  Y(s) 
\non \\[1ex]  
- \dot  Y(s)\,  \mathcal{C}(s)\,   \nabla_L \dr Y(s)  
- \nabla_R \dl Y(s)\, \mathcal{C}(s)\, \dot  Y(s)~. 
} 
The terms on the second line cancel because 
$ \nabla_L \dr Y(s) = \nabla_R \dl Y(s)$ and $\cC^T=-\cC$ is anti--symmetric. Using that 
$  \nabla_a \mathcal{C}_{mn}(s) = 
\der_a Y^p(s)\, \nabla_p \mathcal{C}_{mn}(s)$ the first $s$--derivative of the $\cC$--terms in the action can be rewritten as 
\equa{ 
\frac{\der}{\der s}& 
\Big[ \dl Y(s)\, \mathcal{C}(s) \,  \dr Y(s) \Big] 
 = 
\dl Y(s)\, \nabla_s \mathcal{C}(s)\,  \dr Y(s) 
+ \nabla_L \dot Y(s)\, \mathcal{C}(s)\,  \dr Y(s) 
+ \dl Y(s)\, \mathcal{C}(s)\,  \nabla_R \dot Y(s) 
\non \\[1ex]  &
= \dl Y(s)\, \nabla_s \mathcal{C}(s)\,  \dr Y(s)  
-  \dot  Y(s)\, \nabla_L \mathcal{C}(s)\,  \dr Y(s)  
-  \dl Y(s) \, \nabla_R \mathcal{C}(s)\,  \dot  Y(s) 
\\[2ex]  & 
= \big[ 
\nabla_k \mathcal{C}_{m n} 
- \nabla_m \mathcal{C}_{k n}
- \nabla_n \mathcal{C}_{m k} 
\big](s)\, 
\dot Y^k(s) \dl Y^m(s) \dr Y^n(s) 
= 
\cH_{k m n}(s)\,  \dl Y^m(s) \dr Y^n(s)\, \dot Y^k(s)~. 
\non }
using the definition of the field strength below \eqref{eq:act}. By applying another $s$--derivative we then obtain 
\equa{ 
\frac{\der^2}{\der s^2} 
\Big[ \dl Y(s)\, & \mathcal{C}(s) \, \dr Y(s) \Big] = 
\nabla_{l} \cH_{k mn}(s) \, \dl Y^m(s) \dr Y^n(s) \, \dot Y^k(s)  \dot Y^l(s)
\\[1ex]  
 &+ \cH_{kmn}(s)\, 
\nabla_L  \dot Y^m(s)  \partial_R Y^n(s) \, \dot Y^k(s) 
 + \cH_{kmn}(s)\, 
 \partial_L Y^m(s)  \nabla_R  \dot Y^n(s) \, \dot Y^k(s)~.
 \non 
}
In a similar fashion also the constraint terms can be expanded.

\def\theequation{\thesection.\arabic{equation}} 
\setcounter{equation}{0}
 \section{Aspects of dimensional regularization}
 \label{sc:DimReg} 

We use dimensional regularization to regularize the divergent integrals encounted in this work. For a detailed introduction to dimensional regularization see e.g.\ \cite{tHooft.1972,Collins,Srednicki.2007}. As usual we have introduced the regularization scale, $\gm$, for the integrals to have the same mass dimension as in two worldsheet dimensions when extending to $d = 2 -2\, \epsilon$. 


Define the set of basic integrals 
\equ{ 
I_n(m^2) = \int \dpp\, \frac{1}{(p^2 + m^2)^n} = 
\frac{1}{4\pi}\, \frac{1}{m^{2(n-1)}} \, 
\frac{\Gamma\big(n-\frac{d}{2}\big)}{\gG(n)} \, 
\Big( 4\pi\, \frac{\gm^2}{m^2}\Big)^{1 - \frac d2}~, 
}
depending on some mass parameter $m$. For $n > 1$ these are finite in two dimensions; the fundamental logarithmically divergent integral in two dimensions is:   
\equ{ \label{logdiv} 
I_1(m^2) = \frac 1{4\pi}\, \gG(\ge) \, \Big( 4\pi\, \frac{\gm^2}{m^2}\Big)^\ge = 
\frac{1}{4 \pi}\, \Big[
 \frac 1\ge + \ln \Big( \frac{\bgm^2}{m^2} \Big) 
  \Big]~, 
}
with $\bgm^2 = 4\pi e^{-\gg_E}\gm^2$ where $\gg_E$ is the  Euler--Mascheroni constant. After the last equality we only kept divergent and finite terms. All other divergent integrals can be expressed in terms of $I_1$, for example: 
\equ{
J(m^2) = \int \dpp\,  \frac{p^2}{(p^2 + m^2)^2} = I_1(m^2) - I_2(m^2)~. 
}
Consequently, we have for the tensor valued integral 
\equa{ \label{tensorintegral} 
J_{ab}(m^2) =&\ \int \dpp\,  \frac{p_a p_b}{(p^2 + m^2)^2} 
 \\[1ex] 
=&\  
 \frac 1{d}\, \get_{ab}\, 
\int \dpp\,  \frac{p^2}{(p^2 + m^2)^2} 
 = 
\frac{1}{d} \, \big( I_1(m^2) - I_2(m^2) \big) \, \hat\get_{ab}
= \frac 12\, I_1(m^2)\, \hat \get_{ab} + \text{finite}~, 
\non }
by rotational invariance in $d$ dimensions. Here the generalized worldsheet metric $\hat\get$, that was introduced below \eqref{dextKin}, is used. 


Next we consider integrals that arise in one--loop self--energy diagrams. These integrals depend on an external momentum $k$, e.g.\ 
\equa{\label{momentumtensorintegral}
J_{ab}(k, m^2) =&\ \int \dpp\,  
\frac{p_a p_b}{[(p+\frac 12 k)^2 + m^2][(p-\frac 12 k)^2 + m^2]} 
\\[1ex] 
=&\ \int_0^1 dx\, \int \dpp\,  \frac{p_a p_b + \frac 14 (2x-1)^2 k_a k_b}{(p^2 + M^2(x))^2} 
 = \frac 1{2}\, {I_1(m^2)}\,  \hat\get_{ab} + \text{finite}~.  
\non }
Here we have introduced a Feynman variable $x$ and set $M^2(x) = m^2 + x(1-x)\, k^2$. The integral proportional to $k_ak_b$ is convergent; the momentum integral beeing of the form of \eqref{tensorintegral}. Up to finite contributions the remaining integral over the Feynman parameter is trivial.

\providecommand{\href}[2]{#2}\begingroup\raggedright\endgroup

\end{document}